\documentclass[11pt]{article}
\pdfoutput=1

\usepackage{graphicx}
\usepackage{cite}

\usepackage[in]{fullpage}
\usepackage{caption}
\usepackage{algorithm,algorithmic,amsfonts,amsmath,amssymb,array,booktabs,bbm,bm,color}
\usepackage{dsfont,enumerate,enumitem,epsfig,epstopdf,float,framed}
\usepackage{graphicx,mathtools,multirow,setspace,url,wrapfig}
\usepackage[para]{footmisc}
\usepackage[caption=false,font=normalsize,labelfont=sf,textfont=sf]{subfig}
\usepackage{hyperref}
\usepackage{hhline}
\usepackage{tikz}
\usetikzlibrary{mindmap,trees}

\newcommand{\beq}{\begin{equation}}
\newcommand{\eeq}{\end{equation}}

\setlist[itemize]{noitemsep}

\DeclareMathOperator{\argmin}{argmin}
\usepackage{authblk}

\author[1]{Konstantina Christakopoulou\thanks{christa@cs.umn.edu}}
\author[1]{Arindam Banerjee\thanks{banerjee@cs.umn.edu}}
\affil[1]{Department of Computer Science \& Engineering, University of Minnesota, USA}

\date{}

\begin{document}

\title{Adversarial Recommendation: Attack of the Learned Fake Users}

\maketitle
\begin{abstract}
Can machine learning models for recommendation be easily fooled? While the question has been answered for hand-engineered fake user profiles, it has not been explored for \emph{machine learned} adversarial attacks. This paper attempts to close this gap. 

We propose a framework for generating fake user profiles which, when incorporated in the training of a recommendation system, can achieve an \emph{adversarial intent}, while remaining \emph{indistinguishable} from real user profiles. 
We formulate this procedure as a repeated general-sum game between two players: an oblivious recommendation system $R$ and an adversarial fake user generator $A$ with two goals: (G1) the rating distribution of the fake users needs to be close to the real users, and (G2) some objective $f_A$ encoding the attack intent, such as targeting the top-$K$ recommendation quality of $R$ for a subset of users, needs to be optimized. We propose a learning framework to achieve both goals, and offer extensive experiments considering multiple types of attacks highlighting the vulnerability of recommendation systems.
\end{abstract}

\section{Introduction}
Fake social media accounts are created to promote news articles about a political ideology; false online product reviews attempt to bias users' opinions favorably or against certain products---these are just a few of the many real life examples illustrating that recommendation systems are exposed and can be susceptible to threats from adversarial parties. 

Machine learning algorithms have an ever-growing impact on people's everyday lives.  Recommendation systems heavily rely on such algorithms to help users make their decisions---from which show to watch, to 
which news articles to read (which could end up influencing their beliefs). 
Thus,  a natural question is: \emph{How easy is it to manipulate a machine learned system for malicious purposes?} An answer to such a question would be a stepping stone towards safer artificial intelligence \cite{papernot2016towards,goodfellow2018making}. 

To study this question, 
the first necessary step is the creation of adversarial examples; this would allow one to test the algorithms against them, and potentially increase the algorithms' robustness \cite{goodfellow2014explaining}. With this motivation, a recently thriving subfield of machine learning is the one of adversarial examples---find the minimal perturbation vector to add to the feature vector of an example so that an oblivious classifier misclassifies the perturbed example. These works focus on classification   \cite{szegedy2013intriguing,goodfellow2014explaining,moosavi2016deepfool,moosavi2016universal,su2017one}. 

In \emph{recommendation systems}, the adversarial attacks have a different form. Instead of minimally perturbing an existing example to misclassify it, the attack consists of creating a few adversarial user profiles rating items with some intent.  The intent could be to promote a specific item, or to deteriorate the recommendation quality of a group of users. The setting is not new; in fact, it has been researched since \cite{o2002promoting,lam2004shilling}. However, the injected fake user profiles are hand-coded---typically, the fake users rate the target item with a small or large score, and the rest with random or normal distributed scores to mimic the true rating distribution. 


Our goal is to revisit the question of crafting adversarial fake user profiles for a recommendation system \emph{from an optimization perspective}. 
We pose this as finding a matrix of fake users$\times$items, so that (G1) the distance between the rating distributions of real and fake users is small, and (G2) the adversary's intent is accomplished. 
The scenario is highly realistic---e.g., an adversary $A$ creates a small number of realistic-looking fake user accounts with the goal of removing a target group of a competitor company's products from target users' top lists. We assume that $A$ knows the recommender's model and algorithm to fit the model, and can fit similar models on any new/fake data.

Particularly, we make the following contributions:
\begin{enumerate}
\item We \textbf{formulate} \emph{adversarial recommendation} as a game of an adversary vs. an oblivious recommender, e.g. a low-rank model. There are two objectives: (1) given the real and some fake ratings, learn the low-rank model based on the \emph{recommender's objective} and (2) use the low-rank model to evaluate the \emph{adversarial objective}. This two-step process makes the adversary's task more involved.
\item We propose a \textbf{learning framework} for adversarial attacks on recommendation systems, using: (i) generative adversarial nets (GANs) \cite{goodfellow2014generative} to {learn initial fake users that mimic the true rating distribution} and (ii) suitably update them optimizing an objective encoding the adversarial goal. For (ii), we use \emph{0-th order optimization} to construct the gradient, as the adversary does not have direct access to the gradient. Our framework is the first to find \emph{machine learned} attacks on recommendation systems, allowing to optimize complex intents.  
\item Our \textbf{real-world experiments} show that machine learned adversarial attacks with a wide range of intents are very much possible. As a striking example of a malicious attack we illustrate that in order to  ruin the predicted scores of a specific item for users who would have loved or hated that item, it suffices to minimize the predicted score of the user with the highest predicted score before the attack. 
\end{enumerate}

The rest of the paper is organized as follows. 
In Section~\ref{sec:adv-framework} we formalize the problem of adversarial recommendation and in Section~\ref{sec:adv-method} we propose our learning procedure from the perspective of an adversary of the recommender. We empirically evaluate the proposed methods in Section~\ref{sec:my-adv-experiments}, review related work in Section~\ref{sec:adv-related}, and give a summary in Section~\ref{sec:adv-concl}.

\label{sec:adv-intro}

\section{Problem Formulation}
Our considered 
model for attacking a recommendation system involves two
players: an {\em oblivious} recommendation system $R$ and an {\em adversarial} `fake user' generator $A$. The goal of the recommendation system $R$ is to build a model with parameters $\theta_R$
to minimize a suitable loss function between true and model predicted ratings over all users and items.
The goal of the adversary $A$ is to generate fake users using a model with parameters
$\theta_A$ such that: 
\begin{enumerate}[leftmargin=1cm]
\item[(G1)] the fake users are indistinguishable 
from the real users based on reasonable metrics, e.g., ratings distributions of the fake users are similar to real users,
eigen-spectrum of the fake user ratings are similar to that of real user ratings, etc., and 
\item[(G2)] a recommendation model learned by $R$ using the fake users generated by $A$ leads to worse predicted
ratings for a suitable subset of the real users and/or items, e.g., makes an item less desirable
to a subset of users.
\end{enumerate}


\noindent Let $\mathcal{I}$ be the set of items, and $\mathcal{U}$ the set of real users present in the recommendation system. Let $m = |\mathcal{I}|$ be the number of items and $n = |\mathcal{U}|$ the  number of real users, where $|\cdot|$ denotes the cardinality of a set. Let $X \in \mathbf{R}^{|\mathcal{U}| \times |\mathcal{I}|}$ denote the matrix of ratings from real users.
We assume that the adversary $A$ has a certain budget of $k$ fake user profiles, where $k \ll |\mathcal{U}|$; and that each user profile is a $|\mathcal{I}|$ dimensional vector; aka how the user has rated the different items in $\mathcal{I}$, with zero values denoting empty ratings. Particularly, $A$ outputs a matrix $Z \in \mathbf{R}^{k \times |\mathcal{I}|}$,
where each row $\mathbf{z}_{i'}$ for $\{i'\}_1^k$ is a fake user profile. The total of real and fake users is $n' = n + k$.


The setting can be formulated as a repeated, general-sum game between two players: the \emph{row} player, 
the recommender $R$ and the \emph{column} player, the adversary $A$.
The recommender $R$ maps $(u, j, r)$ tuples to some real-valued score representing the predicted rating of user $u$ on item $j$, and is parameterized by $\theta_R$. The actions of $R$ include all $\theta_R$, e.g., for low rank recommender models, each $\theta_R$
 corresponds to a pair of $U, V$ latent factor matrices. The actions of the adversary $A$ include all fake user profiles $Z$, which are generated using a model parameterized by $\theta_A$.


Both players consider a loss function (the negative of a payoff function) they wish to minimize. If the row player chooses actions $\tilde{U}, \tilde{V}$ (latent factor matrices) and the column player chooses action $\tilde{Z}$ (fake user matrix), then for the row player, the  functional form of the payoff is $f_R (\tilde{U}, \tilde{V}, \tilde{Z})$, and for the column player, is $f_A (\tilde{U}, \tilde{V}, \tilde{Z})$, where the arguments of $f_A(\cdot)$, $f_R(\cdot)$ are the actions played in this round.

In the general setting, each player maintains a distribution over respective action spaces, and will play by drawing
an action from the distribution. 
The row player maintains a distribution $P_r$ over the space of ($U, V$), i.e., $(\tilde{U}, \tilde{V})\sim P_r (U, V)$, and the column player maintains a distribution $P_c$ over $Z$, i.e., $\tilde{Z} \sim P_c(Z)$. 
The distributions of the recommender and the adversary are parameterized by $\theta_R$ and $\theta_A$ respectively. 
In a repeated game setting, let $(\theta_R^t,\theta_A^t)$ be the current parameterizations of the two players.
In the next step, the goal of each player is to find optimal parameters $\theta_R^{t+1}$ and $\theta_A^{t+1}$ 
respectively such that their corresponding expected loss is minimized:
\begin{equation*}
\theta_R^{t+1} = \underset{\theta_R}{\argmin} f_R (\theta_R, \theta_A^t),~~ \theta_A^{t+1} = \underset{\theta_A}{\argmin} f_A (\theta_R^t, \theta_A)
\end{equation*}
or 
\begin{align*}
\theta_R^{(t+1)} & =  \mathbf{E}_{(U,V) \sim P_r(\theta_R), Z \sim P_c(\theta_A^t)} \left[ f_R(U,V,Z) \right] \\
\theta_A^{(t+1)} &= \mathbf{E}_{(U,V) \sim P_r(\theta_R^t), Z \sim P_c(\theta_A)} \left[ f_A(U,V,Z) \right]~.
\end{align*}
Note that $f_R \neq - f_A$, so the game is not zero sum.

We assume that the adversary $A$ knows how the oblivious recommender $R$ fits the model.
In particular, $A$ knows $R$'s loss function $f_R(\cdot,\cdot)$, the parametric representation $\theta_R$, e.g., low-rank
model with latent factors $(U,V)$. Thus, the adversary $A$ can evaluate how predicted ratings will change
for any given fake user ratings matrix $Z$ augmented to the true ratings matrix $X$. However, one main challenge for $A$ is that there is a two-step process going on: (step 1) given some fake ratings, learning say the low-rank model based on the recommendation system objective, typically using non-convex optimization, and (step 2) use the low-rank model to evaluate the adversarial objective. As a result, the adversary typically cannot compute the gradient of the effect
w.r.t. $Z$. 
In the sequel
we approach the problem of constructing $Z$ from the adversary's perspective. 


%
%
%

%
%
%

\label{sec:adv-framework}

\section{Learning}
\label{sec:adv-method}

We detail the specifics of the recommender and adversary considered, and discuss our proposed learning approach.
\\\\
\noindent \textbf{Recommender Strategy.}
We assume throughout that the recommender $R$ is \emph{oblivious} to the existence of an adversary, hence, it optimizes its loss over \emph{all} given data---before the attack over only the $C_{\text{real}}$ original training user-item-rating tuples $\{u_c, j_c, y_c\}_{c=1}^{C_{\text{real}}}$; after the attack over both $\{u_c, j_c, y_c\}_{c=1}^{C_{\text{real}}}$ and the $C_{\text{fake}}$ non-zero ratings of the $k$ fake user profiles, succinctly represented as a sparse matrix $Z \in \mathbf{R}^{k \times m}$ produced by the adversary$A$, resulting in an augmented training set of $\{u_c, j_c, y_c\}_{c=1}^{C_{\text{all}}}$, with $C_{\text{all}} = C_{\text{real}} + C_{\text{fake}}$. In particular, ${R}$, using parameters $\theta_R$ and a goodness-of-fit loss function $\ell(\cdot)$, maps input tuples $\{u_c, j_c, y_c\}_{c=1}^{C^{\text{all}}}$ to estimated scores $\{\hat{y}_c\}_{c=1}^{C_{\text{all}}}$, so that the loss $f_R$ is minimized
\begin{equation*}
\min_{\theta_R} f_{R}(\theta_R, \theta_A) = \min_{\theta_R} \frac{1}{C^{\text{all}}}\sum_{c=1}^{C_{\text{all}}} \ell (y_c, \hat{y}_c(u_c, j_c; \theta_R, \theta_A)).
\end{equation*}

We assume that the recommender is a low rank model; however, our overall approach is not specific to such a model. The low rank recommender has latent factors  $U \in \mathbf{R}^{n' \times d} $ capturing the latent preferences of users, and $V \in \mathbf{R}^{m \times d}$ capturing the latent attributes of the items, and optimizes its expected loss over its parameters $U, V$:
\begin{equation}
\label{eq:UV}
(U^*, V^*)=\arg\min_{U, V} \| [X; Z]  - UV^T \|_2^2 + \lambda \|U\|_2^2 + \lambda \|V\|_2^2
\end{equation}
where $[; ]$ denotes concatenation of two matrices over the row axis. 
We optimize the loss by alternative minimization (alt-min for short), i.e., alternating the closed-form $U, V$ update equations, for a few iterations \cite{mnih2008probabilistic}.
%
%
The model has a probabilistic interpretation: the prior parameter distributions are $\mathbf{u}_i \sim \mathcal{N}(0, \lambda I), \mathbf{v}_j \sim \mathcal{N}(0, \lambda I)$ and conditional
model $X(i, j) \sim \mathcal{N}(\mathbf{u}_i^T \mathbf{v}_j, \sigma)$, where $I$ is the $d\times d$ identity
matrix, $\mathbf{u}_i$ is the $i$-th row of $U$, and $\mathbf{v}_j$ the $j$-th row of $V$. Thus, from a Bayesian perspective, $R$ can maintain a posterior distribution over its parameters $(U, V)$.
For computational simplicity, $R$ is assumed to pick the mode $(U^*,V^*)$ of the posterior distribution. 

The adversary $A$ is aware of the model and algorithm, and is able to compute the point estimates $(U^*,V^*)$ for
any chosen $Z$. Note that since 
an alt-min
algorithm is needed to obtain $(U^*,V^*)$, the adversary does not have a direct way to do gradient descent
w.r.t.~$Z$ on functions of $(U^*,V^*)$; we will return to this point later.\\\\
 \noindent \textbf{Adversary Strategy.}
The adversary $A$, with parameters $\theta_A$, learns and outputs a (distribution over) 
fake user matrix $Z \in \mathbf{R}^{k \times m}$, which should satisfy the two goals presented earlier---(G1) the unnoticeability goal and (G2)  satisfying an adversarial intent. The intent is captured by the loss function $f_A(\theta_R^t,\theta_A)$; 
our experiments explore various intents.

%
%
The intent can be defined over a set of target items $\mathcal{I}_{\text{H}}$, target users $\mathcal{U}_{\text{H}}$, a single target user $u$, or target item $h$. 
Some examples of intents are:
\begin{itemize}
\item {Target the predicted score for ($u$, $h$):}
$f_A = \hat{y}{(u, h)}$
\item {Target the mean predicted score for $h$ over the target users $\mathcal{U}_{\text{H}}$:}\\
$f_{A} = \frac{1}{|\mathcal{U}_{\text{H}}|}\sum_{u \in \mathcal{U}_{\text{H}}} \hat{y}{(u, h)}$
\item {Target recommendation metric@top for $\mathcal{U}_{\text{H}}$:} \\  e.g. Hit Rate (HR), 
$f_A = \frac{1}{|\mathcal{U}_{\text{H}}|} \sum_{u \in \mathcal{U}_{\text{H}}} \text{HR}(u)$.
\end{itemize}

\noindent \textbf{Approach for (G1).} We generate fake users by using generative adversarial nets (GANs) \cite{goodfellow2014generative}. In GANs, a pair of Generator-Discriminator networks pitty each other---the generator $G$ generating samples with the goal of fooling the discriminator $D$ to not being able to distinguish them from real. 
At convergence, the conditional distribution of the generator $G$ 
should give fake user samples which cannot be distinguished by $D$ from real ones. We discuss details of the GANs architecture to generate fake users for recommendation systems in Section \ref{subsec:gans}.\\\\
\noindent \textbf{Approach for (G2).}
To accomplish both (G1) and (G2), one could in principle change the loss of GANs to be a convex combination of two losses: the ``perception" loss of fooling $D$ (as in GANs formulation) and the ``adversarial loss" encoding the adversary's intent, i.e., $f_A$. 

Instead, we opt for a simpler two-step approach: first train GANs until convergence  and sample a set of fake users $Z_1=Z_{\text{GAN}}$ from the conditional posterior of $G$;  and second, suitably modify the sampled users using a variant of gradient descent to optimize
the adversary's intent $f_A$ over the fake users $Z$. In the process, we want to make sure that the resulting fake users helping with
the adversary's intent do not come across as obviously fake. 

Let us consider the problem of interest to the adversary: 
%
\begin{equation}
\label{eq:adv-z}
\min_Z f_A (Z), 
\end{equation}
where recall that $Z$ plays the role of the $\theta_A$ actions of the adversary and the argument $\theta_R^t$, i.e., $U^t, V^t$, is dropped for brevity. 
To optimize \eqref{eq:adv-z} we use \emph{projected gradient descent} for $\{t\}_{1}^{T}$
\begin{equation}
\label{eq:project}
\tilde{Z}_{t+1} = Z_{t} - \eta \nabla_{Z_{t}} f_A (Z; \theta_A), ~ Z_{t+1} = {\Pi}_{\text{allowed range}}(\tilde{Z}_{t+1}) 
\end{equation}
where the projection $\Pi$ is to ensure that the marginals of real and fake users remain close after the descent, $\eta$ is the learning rate, and $\nabla_{Z_{t}} f_A$ is the gradient of the adversarial loss w.r.t $Z_t$. 

Now the question is, how can we compute the gradient $\nabla_{Z_{t}} f_A$? To make things concrete, let us consider as 
adversarial intent: minimize the predicted rating of item $h$ over all {real} users who have not rated the item
\begin{equation}
\label{eq:example}
\min_{Z} \frac{1}{|\{u \notin \text{Ra}(h)\}|} \sum_{u \notin \text{Ra}(h)} \mathbf{u}_u^T \mathbf{v}_h
\end{equation}
At first glance, from \eqref{eq:example} the loss is a function of the recommender's parameters, and not $Z$. But, recall from \eqref{eq:UV} that the parameters of the recommender \emph{are} a function of $Z$---more generally, $f_A$ is a function of ($\theta_A, \theta_R^t$). After playing fake matrix $Z$, the adversary player gets to observe the loss \emph{only for this single $Z$ played}, and not the other actions/ matrices it could have played. Thus, the adversary gets limited information, or else \emph{bandit feedback}. Put differently: the gradient of the loss is not directly given for the optimization over $Z$.

To obtain an \emph{approximation of the gradient}, we build upon 0th-order optimization works in bandit optimization \cite{agarwal2010optimal,duchi2015optimal}. The idea is that if we can only perform query evaluations, to obtain the gradient of $f(Z)$, we need to query $f(Z)$ at two nearby points: $Z_t$ and $Z_t + \alpha Z_0$, for a small $\alpha$ and a suitable fixed matrix $Z_0$. Then we can compute the gradient as the directional derivative along the direction $Z_0$:
$\nabla f(Z_t) = (f(Z_t + \alpha Z_0) - f(Z_t))Z_0/\alpha .$

We use a refinement of Algorithm 3 in \cite{agarwal2010optimal}: there instead of two-point evaluation, they needed $K$ directions and computed the gradient using all $K$ directions. Instead we use as $K$ directions the $K$ top left and right singular vectors of the fake user matrix at round $t$ $Z_t$, obtained from a Singular Value Decomposition on $Z_t$: $Z_t = \tilde{U} \Sigma \tilde{V}^T$.
Let $Z^{(h)}$ be the rank one matrices built from each left and right singular vectors of $Z_t$, $Z^{(h)} = \tilde{u}_h \tilde{v}_h^T$ for $\{h\}_{1}^{K}$, where $K$ is the rank of $Z_t$. Then, we can use these rank-1 matrices $Z^{(h)}$ as $K$ possible directions, and compute the matrix gradient based on these: 
\begin{equation}
\label{eq:approxgrad}
\nabla f(Z_t) = 1/\alpha \sum_{h=1}^K ( f(Z_t + \alpha Z^{(h)}) - f(Z_t)) Z^{(h)}
\end{equation}
This involves $K+1$ evaluations of the function $f(Z)$. To make things faster, we use warm-start---we first evaluate $f(Z_t)$, then, for any $f(Z_t +\alpha Z^{(h)})$, we use the final $(U,V)$ for $f(Z_t)$ to warm start the iterates. 
Similar strategies  
have been studied in stochastic and evolutionary optimization \cite{bhatnagar2012stochastic,hansen2001completely}. 

Algorithm \ref{alg:learning-adv} summarizes our proposed learning approach. 

 \begin{algorithm}
    \caption{Learning Algorithm for Adversary's Strategy}
\label{alg:learning-adv}
  \begin{algorithmic}[1]
  \STATE $U^0, V^0 \leftarrow$ train low rank $R$ over real data $X$.
  \STATE $Z^0 \leftarrow$ train generative adversarial nets on the dataset.
  \FOR{ $t=1, \ldots, T$} 
  \STATE Update the adversary's $Z$ descending its gradient \eqref{eq:approxgrad}.
\FOR{each of the $K+1$ evaluations  $f_A (Z)$}
  \STATE Update the recommender's ($U, V$) with alt-min \eqref{eq:UV}.
  \ENDFOR
\ENDFOR
  \end{algorithmic} 
\end{algorithm}



\section{Experiments}
\label{sec:my-adv-experiments}

We design our experiments to understand the effectiveness of the proposed approach in creating an adversary model $A$ that produces fake users which  cannot be distinguished from real, and which influence in some way the recommender $R$. 

\subsection{Can We Learn Realistic User Profiles?}
\label{subsec:gans}

\begin{figure*}[t]
\centering
\subfloat[Real User Sample (first 64 users)]{
\includegraphics[width=0.45\textwidth]{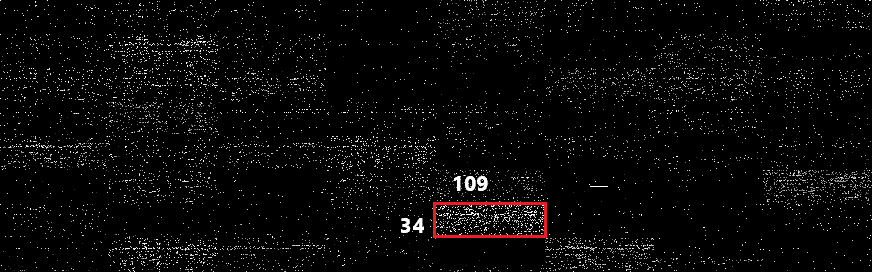}} ~~   
\subfloat[Epoch 0]{\includegraphics[width=0.45\textwidth]{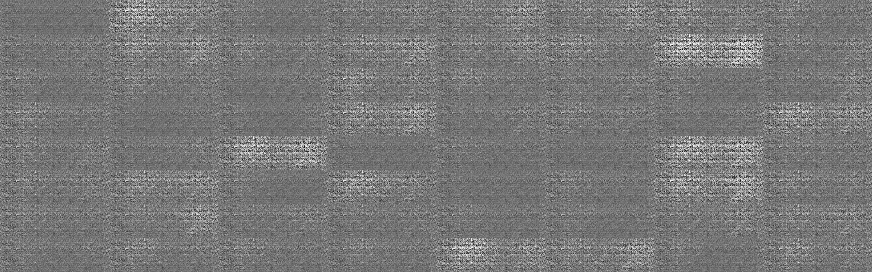}} \\
\subfloat[Epoch 20]{\includegraphics[width=0.45\textwidth]{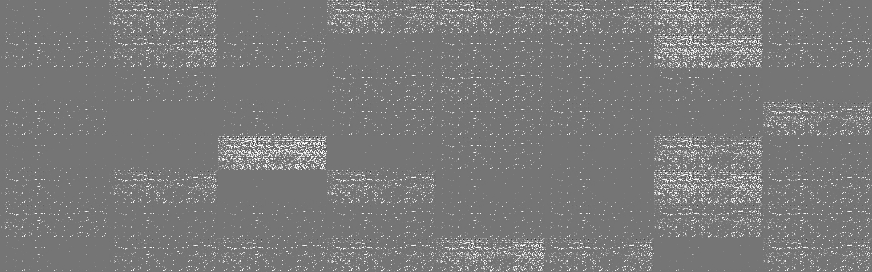}}  ~~
\subfloat[Epoch 90]{\includegraphics[width=0.45\textwidth]{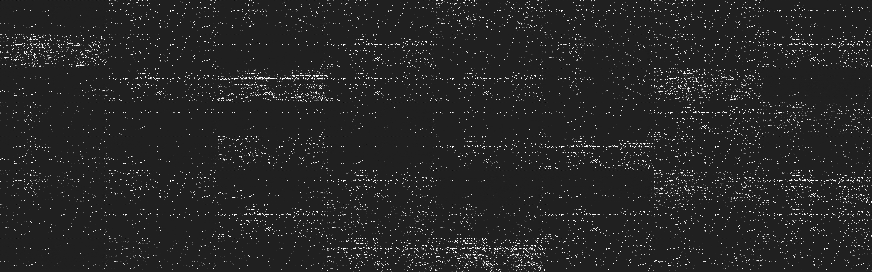}}  
\caption
[Visualization of 64 sampled fake users for MovieLens 1M in a $8\times 8$ grid. As the training epochs progress, $G$ learns to generate realistic-looking fake users. ] 
{
Visualization of 64 sampled fake users for MovieLens 1M in a $8\times 8$ grid. 
The red box in (a) surrounds one real user sample, which is of dimension 34$\times$109---when flattened it is $3706$-dimensional, i.e., $|\mathcal{I}|$. As the training epochs progress, the generator network $G$ of GANs learns to generate realistic-looking fake users.}
\label{fig:ML1M}
\end{figure*}

The first question we investigate is whether with generative adversarial nets we can learn fake user profiles that seem like real. \\\\
\noindent \textbf{Network Architecture.} We used the DCGAN architecture \cite{radford2015unsupervised}, thanks to its good empirical performance. The Discriminator $D$ takes an image of size $H \times W$ (fake or real user sample) and outputs either a $0$ or a $1$ (is it fake or real?). It consists of four 2D convolutional (\texttt{CONV}) units, with \texttt{leaky ReLUs} and batch normalization (\texttt{BN}), whose depths are respectively $[64, 128, 256, 512]$, followed by a single-output fully connected (\texttt{FC}) unit with sigmoid activation. The Generator $G$ takes as input noise $z \sim \mathcal{N}(0, 100)$ and outputs an $H \times W$ image (the fake user sample). It consists of a \texttt{FC} unit of dimension $2\times 4 \text{~(or 7)} \times 512$ with \texttt{ReLU} and \texttt{BN}, reshaped to a  $2, 4 \text{~or 7}, 512$ image, followed by four \texttt{transposed CONV} units of depths $[256, 128, 64, 1]$ respectively, each with \texttt{ReLU}  and \texttt{BN}, except for the final with a tanh. 
We set for the (transposed) \texttt{CONV} units the stride to 2, the kernel size to $5 \times 5$. 
 We set batch size to 64, and run DCGAN for 100 epochs (each epoch does a cyclic pass over all batches.)\\\\
\begin{table}[!t]
\centering
\begin{tabular}{|c|c|c|c|}
\hline
Dataset & \# of Items & 2D Shape & \# of Users \\
\hline 
MovieLens 100K & 1682 &  $29 \times 58$ & 943\\ 
MovieLens 1M & 3706 &  $34 \times 109$ & 6040\\ 
\hline 
\end{tabular}
\caption{Dataset Statistics. The 2D Shape is the conversion of the $|\mathcal{I}|$-d vector to a 2-D H $\times$ W ``image". 
}
\label{tab:1}
\end{table}
\noindent \textbf{Datasets.} 
Since we want to learn user profiles for recommendation systems, we used two popular movie recommendation datasets, MovieLens 100K, MovieLens 1M \cite{harper2016movielens}, whose statistics are shown in Table \ref{tab:1}. They  contain the ratings of users on different movies in the scale $\{0, 1, 2, 3, 4, 5\}$ with 5 the highest like, and 0 denoting that the user has not rated the movie. As the last layer of $G$ has the tanh activation function, fake user samples are in $[-1, 1]$; hence, using $r' = (r - 2.5)/2.5$ we transformed the real ratings from $[0, 5]$ to $[-1,1]$. \\\\ 
\noindent \textbf{Setup.} Each user profile is an $|\mathcal{I}|$-d sparse vector, which needs to be transformed to a  $H \times W$ 2D array to go through the DCGAN 2D (de-)convolutional units. We set as $H$ the smallest factor of $|\mathcal{I}|$  
and as $W = |\mathcal{I}| / H$. This way, each user is viewed as a 2D  $H \times W$ image with pixel values the ratings of the user on the different items.  \\\\ 
\noindent \textbf{Results.} 
Periodically during training, we sample 64 fake users from the conditional posterior of the generator $G$ and visualize them in a grid of 8 by 8 H$\times$W images. 
In Figure \ref{fig:ML1M}
we illustrate the progress of the samples during training  for MovieLens 1M, with the first column visualizing the first 64 real users. We see that in the first epochs the sampled users are noise, but as training goes on, the real distribution seems to be learned; similar results hold for MovieLens 100K. 


We also want to validate quantitatively that  the fake user distribution is close to the real one at DCGAN's training convergence. For these experiments, we sample 700 fake users from $G$ so that the size of the real and fake user distribution---at least for MovieLens 100K---is comparable. 

We compute the correlation matrix over items of the fake data $Z^{T}Z$ (based on $Z$ sampled from the conditional posterior of the learned $G$ of the last training epoch), and the correlation matrix of the real data $X^{T}X$, and we compare their respective eigenspectrum; specifically their \textbf{top-10 eigenvalues}. 




We compute \textbf{distance metrics} between the real and fake user distributions: 
For each item $j$, the real users form a distribution $P^j$ over the rating values $[-1.0, 1.0]$ (i.e., for MovieLens 100K with 943 samples), and the fake users $G(z)$ form a $Q^j$ distribution over $[-1.0, 1.0]$ (with 700 samples). 
Then, we discretize the values to the six bins $[-1.0, -0.6, -0.2, 0.2, 0.6, 1.0]$ (corresponding to $[0, 1, 2, 3, 4, 5]$): for each of the six bins, we compute the fraction of (real or fake) users who have rated $j$ in this bin out of all (real or fake) users. For a certain item $j$, we compute two such six-dimensional vectors, one for the real users, and a second for the fake users, and then use the following metrics to measure distance between distributions \cite{goodfellow2016deep}:
\begin{align}
& \textbf{{Total Variation Distance (TVD)}} (P^j, Q^j) = \sum |P^j-Q^j|/2 \label{eq:tvd} \\
& \textbf{{Jensen-Shannon Divergence (JS)}} (P^j, Q^j) = \frac{1}{2} (D(P^j || M^j ) + D(Q^j || M^j ))
\label{eq:kl}
\end{align} where $M^j = \frac{1}{2}(P^j + Q^j )$ and $D$ is the Kullback-Leibler divergence. After we have computed the distance metric for each item, we report the average over all  items, i.e, mean TVD and mean JS Div. 


\begin{figure*}[t]
\centering
\subfloat[Top-10 Eigenvalues]{
\rotatebox{90}{\qquad\tiny{\bf Eigenvalues / Index}}
\includegraphics[width=0.4\textwidth]{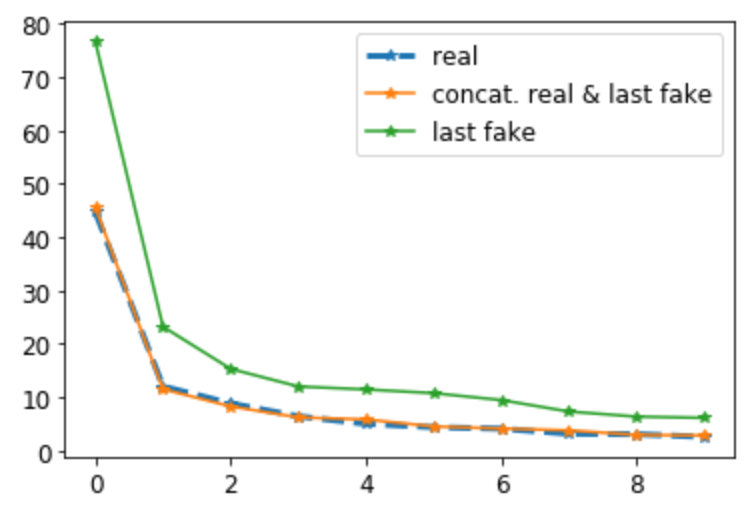}
}
\qquad
\subfloat[Jensen-Shannon (JS) Divergence]{
\rotatebox{90}{~~~\tiny{\bf JS Div. / \# Train. Epochs}}
\includegraphics[width=0.4\textwidth]{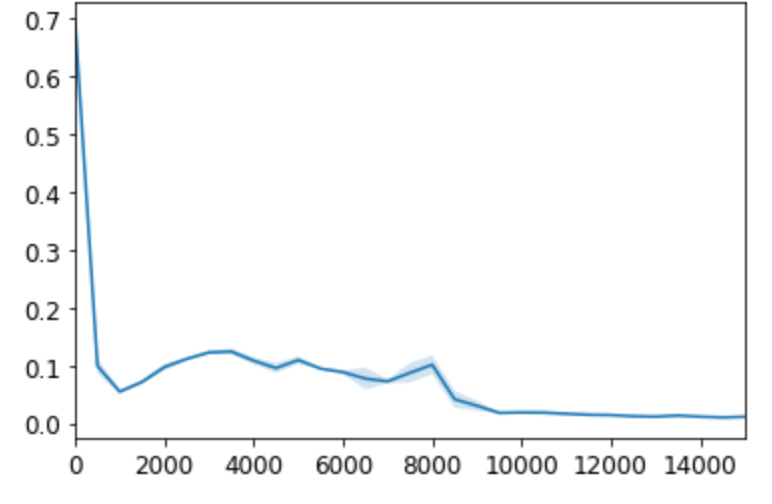}
}~~
\caption{
(a) x-axis: index of top eigenvalue, ranging from 0 to 9, y-axis: corresponding eigenvalue. The eigenspectrums of the correlation matrices of the real, the fake, and the [real; fake] concatenated data are similar. (b) x-axis: DCGAN training iteration, y-axis: mean JS. At convergence, the distance between the real and fake distributions is approximately $ 0$. 
}
\label{fig:gan-quantitative-ML1M}
\end{figure*}
In Figure \ref{fig:gan-quantitative-ML1M} we plot the (a) top-10 eigenvalues and (b) JS Divergence for the MovieLens 1M dataset. 
For (b), the reported results are averaged over five different runs of DCGAN.
The results shown, along with similar results we have observed for MovieLens 100K, indicate our first key finding:
\begin{quote}
{ \emph{Generative adversarial nets can produce fake user samples whose distribution is close to the real user distribution.}}
\end{quote}

\subsection{Experimental Design}
\label{subsec:exp-design}
To evaluate the adversary $A$'s capability of attacking a recommender $R$, we use the two-phased approach described in Section \ref{sec:adv-method}: (1) We first train DCGAN on the respective dataset, as presented before in \ref{subsec:gans}, and (2) we then perform the $Z$-SGD updates, which are initialized by a sample of DCGAN-generated fake users, denoted by $Z_{\text{GAN}}$, transformed from the $[-1, 1]$ to the expected by the recommender range of $[0, 5]$. 

Throughout all experiments, the sample size of $Z_{\text{GAN}}$ will be set to 64; the 64 fake users, iteratively optimized during the $Z$-SGD updates, represent only 0.063 fraction of all system users (real and fake) for MovieLens 100K, and 0.01 fraction for MovieLens 1M.
 
We use two types of experimental setups:
\begin{itemize}
\item[(E1)] $A$ targets unrated user-item entries (thus entries which are candidates for recommendation) not included in the training of $R$. \item[(E2)] $A$ targets a small subset from the recommender's true (user, item, rating) tuples, which is held out from the training of $R$.  
\end{itemize}
For (E2), we define a target set where the adversarial loss $f_A$ is optimized over, and a test set which is unknown to both the recommender and the adversary---the test set is used to check whether the success of adversarial intent generalizes from the target to the test set. We form the target and test sets in two ways: (E2-a) either using the leave-one-out setup, i.e., leaving one tuple per user in the target set (and another tuple in the test set), or (E2-b) an 80-10-10 split, i.e., splitting the original dataset into 80\% of the total ratings per user for training, 10\% for the target set, and 10\% for the test set.


The recommender $R$ under attack is a matrix factorization (else low rank) model, trained on explicit ratings in the scale $\{0, 1, 2, 3, 4, 5\}$. 
Unless otherwise specified, we set the {latent factor dimension $d$} to 40, the {regularization parameter $\lambda$} to 0.001, and $R$ is trained before the attack for 10 {alt-min iterations}. 

For the adversary $A$, we set the SVD approximation rank $K$ to 30, and the {approximate gradient step constant} $\alpha$ to 0.0001.  During a single Z-SGD iteration for each of the $K+1$ $f_A$ evaluations, 5 alt-min iterations of $R$ are performed. 


 We perform warm-start, i.e., for the $t+1$ $Z$-SGD iteration, $R$'s parameters are initialized from the ones obtained at the end of the alt-min $R$ iterations from the previous $t$ $Z$-SGD step. 

To be consistent with the original movie ratings from the datasets, every time $f_A$ is evaluated, e.g. either during the approximate gradient computation or the loss computation, the $Z$ values are rounded to the closest integers, and get clipped to $[0, 5]$. Also, to ensure that while performing the $Z$-SGD updates, the fake user distribution does not diverge from the real distribution, we perform \emph{projected} gradient descent: 
\begin{align}
\tilde{Z}_{t+1} &= Z_t - \eta \nabla_{Z_t} f_A \\
Z_{t+1} &= \text{clip}(\tilde{Z}_{t+1}, 0, 5)\label{eq:box-projection}, 
\end{align}
where \eqref{eq:box-projection} corresponds to a box-projection. 

To evaluate the success of the adversary, we use various metrics; every time we introduce a new metric, we will use bold letters. Overall, we use the metric of \textbf{Attack Difference}, (or else \textbf{magnitude of the attack}) denoted by $\Delta (Z)$: 
\begin{equation}
\label{eq:delta_z}
\Delta (Z) = f_\text{before} (X) - f_{A} (X; Z)
\end{equation}
where $f_\text{before} (X)$ denotes the value of the adversarial loss before $Z$ is concatenated with the real users' data (before the attack). In other words this metric  expresses the decrease of the adversarial loss. In order for the attack to be considered successful, $\Delta (Z)$ needs to be at least positive, and ideally larger than zero by a certain margin. Depending on the intent of the adversary, as encoded by $f_A$, this metric could imply for example a decrease in the predicted score of a target item---in which case, it is related to the ``prediction shift metric" \cite{lam2004shilling,o2004collaborative}---, or a change in the recommendation quality of a target group of users. 


Each of the following sections introduces a separate attack type, as specified by target user(s), item(s) and intent of $A$.

\subsection{Targeting a User-Item Pair}
\label{subsec:singleui}
We start with the adversarial intent: can $A$ learn realistic users that reduce the predicted score for an \emph{unrated user-item entry}? For this, we adopt the (E1) experimental setup.

Let target user be denoted with $u$ and target item with $h$, where $h \notin \text{RatedBy}(u)$. The adversarial loss is the predicted score for $(u, h)$: 
$f_A^{(u, h)} (X; Z)= \hat{y}(u, h)$
and the magnitude of the attack is 
$\Delta^{(u, h)} = f_\text{before}^{(u, h)} (X) - f_A^{(u, h)} (X; Z), $
where $f_\text{before}^{(u, h)} (X)$ denotes the predicted score for the pair by R before the attack. 

We set $\eta$ to 100, $\alpha$ to 50, $K$ to 5. The adversary performs a total of $T=21$ $Z$-SGD iterations, or fewer if a certain stopping criterion is satisfied. We explore two stopping criteria, and two cases for how to specify the $(u, h)$ target entry.


\subsection*{Stopping criterion is $\Delta \geq 1$}
First, we considered to stop the $Z$-SGD iterations when  the predicted score for ($u, h$) is decreased by at least 1 after the attack. We performed this experiment for 70 uniformly at random sampled items for the MovieLens 100K dataset. For each target item $h$, we sampled target user $u$ from the set of users who have not rated $h$. Considering an attack successful only if $\Delta $ is larger than 0,  we found that for only 2 out of the 70 sampled target items the attack was not successful.

\begin{figure*}[t]
\centering
\rotatebox{90}{~~~\tiny{\bf Metrics / \# Z-SGD Iterations}}
\includegraphics[width=0.45\textwidth, height=0.3\linewidth]{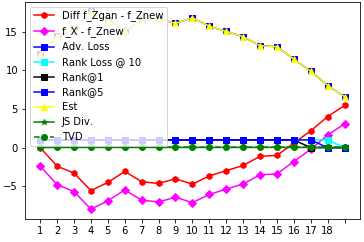}
~~
\rotatebox{90}{~~\tiny{\bf Rank Loss / \# Z-SGD Iterations}}
\includegraphics[width=0.45\textwidth, height=0.3\linewidth]{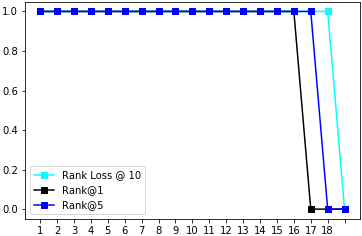}
\caption[The adversary $A$ targets the top-1 item $h$ ID-1062 ``A Little Princess (1995)" for user $u$ ID-0, with stopping criterion that $h$ is removed from the top-10 recommendation list of $u$.]
{The adversary $A$ targets the top-1 item $h$ ID-1062 ``A Little Princess (1995)" for user $u$ ID-0, with stopping criterion that $h$ is removed from the top-10 recommendation list of $u$. x-axis: SGD iterations varying from 1 to 19, with 1 corresponding to using as $Z$ the $Z_{\text{GAN}}$. \textbf{Left:} The y-axis represents metrics capturing the success of $A$ in targeting $(u,h)$, and distance metrics for real-fake user distribution. \textbf{Right:} Zooming in the ranking loss @$top:\{1,5,10\}$. 
\textbf{Gist:} As Z-SGD updates progress, (i) the adversarial loss (`Adv. Loss' $f_A$, blue line), which is the same as the estimated score of $R$ on the target $(u, h)$ pair (`Est.', yellow line) goes down; (ii) the attack difference $\Delta(Z)$ (magenta line) increases, and the difference of the score $f_A(X; Z_{\text{GAN}}) - f_A(X;Z)$ (red line) increases too; (iii)  the rank losses @ top-1/ 5/ 10 (black/ blue/ cyan lines) reach 0 after 17/ 18/ 19  $Z$-SGD iterations; (iv) the real-fake distribution distance metrics (mean `JS Div' mean `TVD', green lines) remain close to 0, despite the adversarial updates.
}  
\label{fig:single-ui-push-example}
\end{figure*}

\subsection*{Targeting Top Item of User, Stopping criterion: Remove from Top}
Next, we used the early-stopping criterion that the item does not exist in the top of the recommendation list anymore, as this aligns better with the actual user experience in a recommendation setting. For this, we randomly sampled target users, and for each user $u$, we considered as target item $h$ the one out of the unrated set predicted to be at the top of the user's list before the attack---simply put, the top item $h$ of user $u$. Beyond the attack difference metric \ref{eq:delta_z}, and the distance metrics  \eqref{eq:tvd}, \eqref{eq:kl}, we report the metric of $\text{\textbf{Rank Loss @ top}}(h)$ = $\mathbbm{1}[\text{item } h \text{ @ top-list}]$, where we considered $top=\{1, 5, 10\}$. 


In the first experiment, for the stopping criterion, we set to remove $h$ from the top-1 list: we found that out of the 135 sampled target users, only for one user the top-one item remained at the top. 

In the second experiment, we set for the stopping criterion to remove $h$ from the top-10. We found that out of the 55 sampled users, only for two users the attack was not successful---for the rest, notably, the adversary managed to remove the target item from the target user's top-10 list, while looking realistic. 
As an example, Figure \ref{fig:single-ui-push-example} illustrates the metrics for the movie ``A Little Princess" that appeared in the top-1 of user ID-0 before the attack. We can see that the attack is successful, removing the top item from the top-1 at iteration 17, and from top-10 at iteration 19 (thus stopping the Z-SGD updates), optimizing well the adversarial loss of the estimated score for the $(u, h)$ entry (yellow line), while the real-fake distribution distance metrics of mean TVD and mean JS Div (green lines) remain close to 0. 



This experiment illustrates our second key finding: 
\begin{quote}
\emph{The adversary can successfully target the top-1 predicted item of a user, and remove it from the top-10.}
\end{quote}

\subsection{Targeting Item's Mean Predicted Score}
\label{subsec:revisit-push-avg-score}

Here, we examine whether the adversary can accomplish a more ambitious goal: can $A$ target (push down) the mean predicted score of a target item $h$ over \emph{all} real users who have not rated $h$ in the training dataset?---again, we adopt the (E1) experimental setup. The reason for choosing this target user set is because these are the users for which $h$ can be a candidate item for recommendation. This intent can be formulated as: 
\begin{equation}
\label{eq:fadv}
f_A^h=  \sum_{u \in \mathcal{U}, \notin \text{Ra}(h)} \hat{y}(u, h),
\end{equation}
and the attack will be successful if $f_{\text{A}}^h (X;Z)$ becomes smaller than the predicted score $f_{\text{before}}^h (X)$. 

\begin{figure*}[!t]
\centering
\rotatebox{90}{~~~\tiny{\bf Metrics / \# Z-SGD Iterations}}
\includegraphics[width=0.45\textwidth]{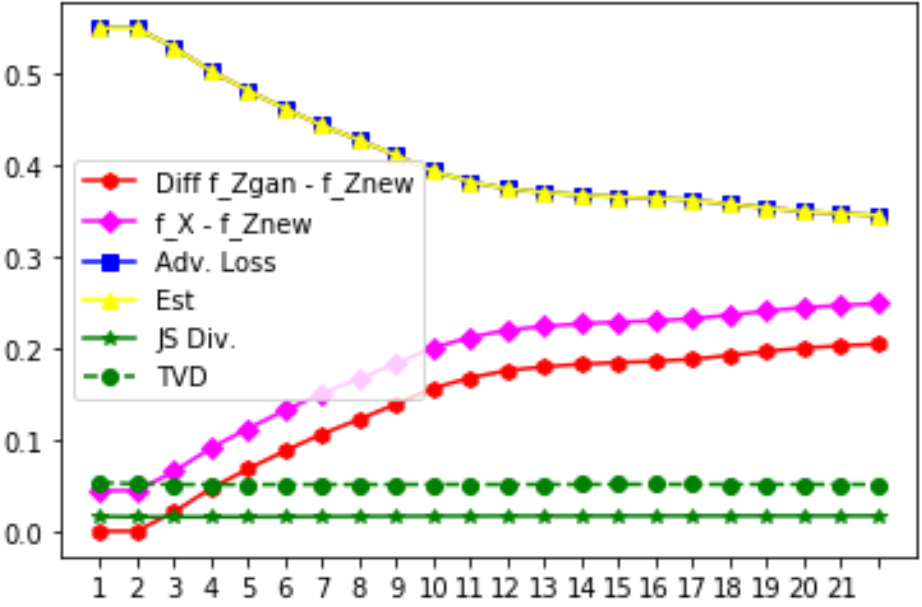}
~~
\rotatebox{90}{~~\tiny{\bf Users' $\Delta$s / \# Z-SGD Iterations}}
\includegraphics[width=0.45\textwidth]{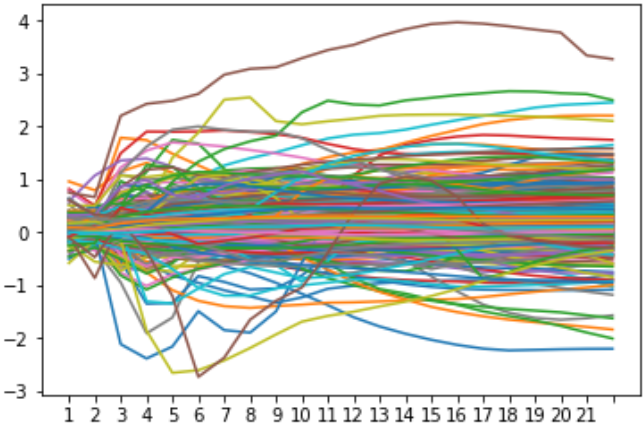}
\caption[The adversary $A$ targets item's $h$ ID-1348 ``Mille bolle blu (1993)" average predicted score by $R$ over all users $\{u \in \mathcal{U}\} \notin \text{Ra}(h)$, with stopping criterion that $\Delta(Z)\geq 1$. ]
{The adversary $A$ targets item's $h$ ID-1348 ``Mille bolle blu (1993)" average predicted score by $R$ over all users $\{u \in \mathcal{U}\} \notin \text{Ra}(h)$, with stopping criterion that $\Delta(Z)\geq 1$. x-axis: SGD iterations varying from 1 to 22, with 1 corresponding to using as $Z$ the $Z_{\text{GAN}}$. \textbf{Left:} The y-axis represents the average metrics over all users. The metrics presented are the same as in Figure \ref{fig:single-ui-push-example}, except for the rank loss metrics which are not included. The `Adv. Loss'/ `Est.' is given by \eqref{eq:fadv}. \textbf{Right:} Every line corresponds to each user's $\Delta(Z)$ varying with Z-SGD updates; the average over these lines is the magenta line on the left plot. \textbf{Gist:} Although on average the attack can seem successful, i.e., the $\Delta(Z)$ increases and the distance metrics remain close to 0 (left), each user's $\Delta$ follows its own trend (right). We conclude that such an attack is difficult.}
\label{fig:distr-users-example}
\end{figure*}

We keep the same setting as before, except for setting $\eta = 1000$ and choosing $\alpha$ in $\{500, 1000\}$, as  we found that for this experiment larger values tend to lead to larger $\Delta$. The early-stopping criterion is $\Delta(Z)$ $\geq 1$. We performed this experiment for 29 randomly chosen target items from MovieLens 100K and found that out of the 29 items, only for 6 items early stopping was realized. Also, for 6 out of the 29 items, the attack was unsuccessful; $\Delta \leq 0$, i.e., the average score after the attack remained the same or increased.  Overall, we conclude that:
\begin{quote}
\emph{Targeting the average predicted score of an item is a hard task.}
\end{quote}

To understand why this happens, we examine how the distribution of $\Delta$ over all users who have not rated target item $h$ evolves over the Z-SGD iterations. From Figure \ref{fig:distr-users-example} we can see for the sampled movie ``Mille bolle blu (1993)"  (similar behavior is noticed in the others too), that although the average difference reached ~0.2 (magenta line in Figure \ref{fig:distr-users-example}, \emph{left panel}), every user's attack difference follows its own trend (Figure \ref{fig:distr-users-example}, \emph{right panel}); with mainly the users with the largest or smallest $\Delta$ affecting the average $\Delta$.
This shows that the fake users cannot move all users' scores on $h$ \emph{simultaneously} to the same direction. 


\subsection{Targeting the Top User of an Item}
\label{subsubsec:impact}

In reality, to attack a target item, the adversary does not need to solve the more difficult problem of pushing down \emph{all unrated users}'s score. Instead, they only need to push the score of users who would be good candidates for getting this item in their recommendations. Put differently, these are the users with the higher predicted scores from $R$ before the attack; the rest of the users would not get $h$ in their recommendations either way.

In this experiment, the adversary's intent is to target the \emph{top user} of an item, i.e., the user from $u \notin \text{Ra}(h)$ with the largest predicted score from $R$ before the attack. This can be seen again as a \emph{targeting a single $(u, h)$ entry attack}, that we found earlier in Section \ref{subsec:singleui} to be a  successful attack; but while $h$ can be any arbitrary target item, $u$ is the \emph{top user} of the item.

\begin{figure}[!t]
\captionsetup[subfigure]{labelformat=empty}
\rotatebox{90}{\parbox{0.3\linewidth}{~~~\tiny{\bf Avg. $\Delta$ / \# Top or Bottom Users}}}
\subfloat[Left panel]{
\includegraphics[width=0.45\textwidth]{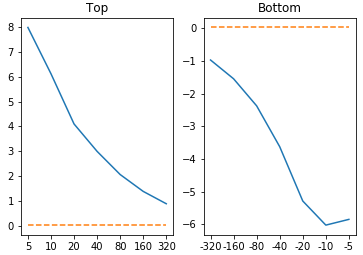}
}
\hfill
\rotatebox{90}{\parbox{0.3\linewidth}{~~~\tiny{\bf Avg. $\Delta$ / \# Top or Bottom Items}}}
\subfloat[Right panel]{
\includegraphics[width=0.45\textwidth]{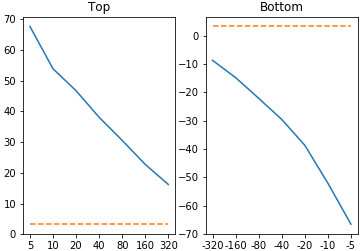}
}
\caption
[The adversary $A$ targets the \emph{top user} $u$ of item $h$ ID-1417 ``The Joy Luck Club" (1993).]
{\textbf{Main Result.} The adversary $A$ targets the \emph{top user} $u$ of item $h$ ID-1417 ``The Joy Luck Club" (1993). We say \emph{top/ bottom} to refer to what was predicted with the highest/ lowest score from $R$ before the attack. We focus on the metric of attack difference $\Delta(Z)$ using the $Z$ obtained after the 21 Z-SGD updates, or when stopping criterion $\Delta(Z)\geq 1$ was satisfied; but we define $\Delta(Z)$ over different subsets of users or items. \textbf{Left:} x-axis: varying number of users considered to compute $\Delta(Z)$, from 5 to 320 top users, or from 320 to 5 bottom users, for item $h$. y-axis: The orange dotted line is the average over \emph{all} users $\notin \text{Ra}(h)$, computed to .48. The blue line connects points of the average $\Delta(Z)$ over the corresponding subset of users, e.g., for the top-5 users $\Delta(Z)=8$, for the bottom-5 users $\Delta(Z)=-6$.  \textbf{Right:} x-axis: varying number of items out of $\notin \text{RatedBy}(u)$ considered to compute $\Delta(Z)$, from 5 to 320 top items, or from 320 to 5 bottom items for user $u$. y-axis: Same as left panel, but instead of top/bottom users for $h$, here $\Delta$ is computed over top/ bottom items for $u$.  
\textbf{Gist:} Targeting the top predicted user of an item attacks also the top-$K$ predicted users of this item, and the top-$K$ predicted items of this user. Also, the bottom predicted users/ items are attacked, as for them the score is increased, which is the opposite from what they would want.  
}
\label{fig:inflonothers-b}
\end{figure}
\begin{figure*}[!t]
\captionsetup[subfigure]{labelformat=empty}
\rotatebox{90}{\parbox{0.4\linewidth}{\tiny{\textbf{ $\Delta(u', h)$ $\forall u' \notin \text{Ra}(h)$ / $u'$-$u$ Correlations}}}}
\subfloat[Left panel]{
\includegraphics[width=0.45\textwidth]{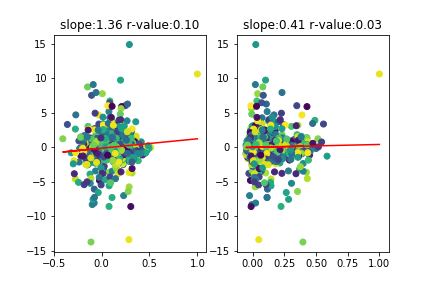}
}
~~~
\rotatebox{90}{\parbox{0.4\linewidth}{\tiny{\textbf{ $\Delta(u, h')$ $\forall h' \notin \text{Ra}(u)$ / $h'$-$h$ Correlations}}}}
\subfloat[Right panel]
{
\includegraphics[width=0.45\textwidth]{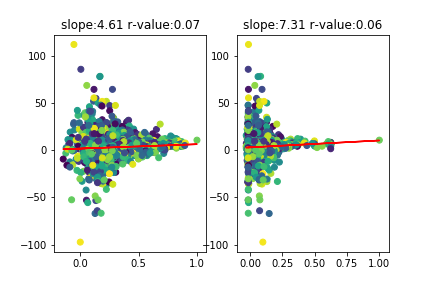}
}
\caption[The effect of targeting the top user $u$ for item $h$ ID-1417 on the rest of users who have not rated $h$; and on the rest of items that have not been rated by the top-user $u$.]
{Same setup as Figure \ref{fig:inflonothers-b}. The effect of targeting the top user $u$ for item $h$ ID-1417 on the rest of users who have not rated $h$; and on the rest of items that have not been rated by the top-user $u$. For both cases, each dot in the scatter plot (colors are for aesthetic purposes) is the attack difference for a single (user, target item) entry or (target user, item) entry. 
\textbf{Left:} x-axis represents the user-target user correlations, computed either based on the estimated latent factors $U$ before the attack (left), or based on the real ratings, i.e., rows of $X$ (right).
\textbf{Right:} x-axis represents the item-target item correlations, computed either based on the estimated latent factors $V$ before the attack (left), or based on the real ratings, i.e., columns of $X$ (right).
\textbf{Gist:}
The red line represents a linear-line fitting on the data points to find if there is a relationship between $\Delta(Z)$ and correlations. Although the line-slope is positive, the $r^2$ is small, thus rejecting the hypothesis for a relationship. However, when only the most correlated ($>.35$ correlation) users are considered, $r^2$  is $.74$ for latent-factor based correlations---showing that 
as the correlations increase, $\Delta(Z)$ also increases. Such a relationship does not hold for the case of items ($r^2=.29$ for all items with item-item correlation $>.6$ ); we see though that as item-target item correlations increase, the variance over the $\Delta$s reduces.
}
\label{fig:inflonothers-a}
\end{figure*}

 In Figures \ref{fig:inflonothers-b}, \ref{fig:inflonothers-a} we show the results of targeting the top user $u$ of item $h$ ``The Joy Luck Club" (similar results hold for other target movies). We can see that just by targeting the predicted score of the most-wanting-``The Joy Luck Club" userID 1417, the scores of all other users who have not rated $h$ get affected too; similarly the scores of userID 1417 for all the other movies he has not rated get affected.


Figure \ref{fig:inflonothers-b} shows how the mean attack difference (y-axis) when considering only the top/ bottom users for the item $h$ (\emph{left panel}), or when considering only the top/ bottom items for the user $u$ (\emph{right panel}), varies as we vary the top/ bottom (x-axis). 

Figure \ref{fig:inflonothers-b}, \emph{left panel} shows that although the average difference over all users who have not rated $h$ is only .48, if we consider only the top-5 users with the highest prediction scores for $h$ before the attack, the average difference is 8. In fact, for the top-80 users the average difference is larger than 2, whereas for the users who were predicted to hate the item the most (bottom-5) the average difference is close to -6; this means that for users who were predicted to least like the item, the predicted score increased, which is opposite to what a good recommender should do. 

The \emph{right panel} illustrates a similar story for the items. The average difference on the items which were predicted to be liked the most by $u$ before the attack is $>$ 60 (items which were a good fit for this user are pushed down), and the average difference on the items which were predicted to be liked the least by the user is ~-70 (items which were a bad fit for the user are pushed up).  

This illustrates an important finding of this work: 
\begin{quote}
\emph{ A successful attack on item $h$ is: When $A$ targets the score of the top user, i.e., the user $u$ predicted by $R$ to like $h$ the most before the attack, then the top-$K$ users for $h$ and top-$K$ items for $u$ are also attacked.}
\end{quote}

On a side note, we want to see whether there is a relationship among the attack differences $\Delta$ of the various users for the target item $h$ with the user-target user $u$ correlations; 
or among the $\Delta$s of the target user $u$ for the various items with the item-target item $h$ correlations.      
Thus, Figure \ref{fig:inflonothers-a} shows how the different users' (left panel) or items' $\Delta$s (right panel) vary as a function of the correlation with the target user or target item respectively. We compute the correlations (a) based on the estimated latent factors $U$ ($V$) of the $R$ before the attack, or (b) based on the true rating matrix $X$. We find that for  items with small correlation with the target item, the variance of the differences is large; as the correlation increases, the variance reduces. We also find that for the users who are most correlated with the target user ($>0.6$ factor-based correlation), as the correlation increases the attack difference becomes larger. 

\subsection{Targeting a Group of Items}
\label{subsec:groupbased}


Next, we focus on attacks that target an entire \emph{group} of items, in contrast to the presented experiments so far, where a single item was the target of each attack. We examine two adversarial goals: 
\begin{itemize}
\item[(A1)] minimize the mean predicted score over all items in a group, and 
\item[(A2)] maximize the prediction error, as measured by mean absolute error, over a group.
\end{itemize}
\emph{Experimental Setup.} We adopt the (E2) experimental setup (Section \ref{subsec:exp-design}) of using the extra information of a ``target set"---the adversary $A$ has the added power of, besides making queries to access $R$'s predictions, being able to target some held-out tuples of (user, item, score) which have not been used as part of $R$'s true training data. This is in contrast to the (E1) setup where $A$ targeted one or a set of \emph{unrated} user-item entries. 

We used the (E2-b) setup of 80-10-10 split of ratings per user.

To define the target item groups, we explored four different ways: (i) grouping them into 10-percentile groups based on the  predicted scores from $R$ before the attack, (ii) using the side information of movie genre, where the same movie can belong to multiple groups, (iii) 10-percentile groups based on the prediction error of $R$ before the attack, or (iv) 10-percentile groups based on number of training ratings per item. To be more precise, for (i) and (iii), we first computed for each item $h$ $R$'s average predicted scores, and average mean absolute error respectively,  over the corresponding (user, $h$, score) tuples in the target set, and we then divided them into deciles based on these values.

\emph{Setting.} For $R$, we set $d$ to 100, and $\lambda$ to 0.1, and we train it for 100 alt-min iterations before the attack. For the adversary $A$, we set $\eta$ to 1000, $K$ to 5, $\alpha$ to 50, and $T=30$. We report the best results from $A$'s side, for the Z-SGD iteration with the best value of $f_A$ in the target set.  
We report the 
\beq
\textbf{\% Target Improved}  = \frac{\Delta * 100}{f_{A}^{\text{before}}},
\eeq where $\Delta$ is given by \eqref{eq:delta_z}. 
 


\begin{figure*}[t]
\centering
\includegraphics[width=0.45\textwidth]{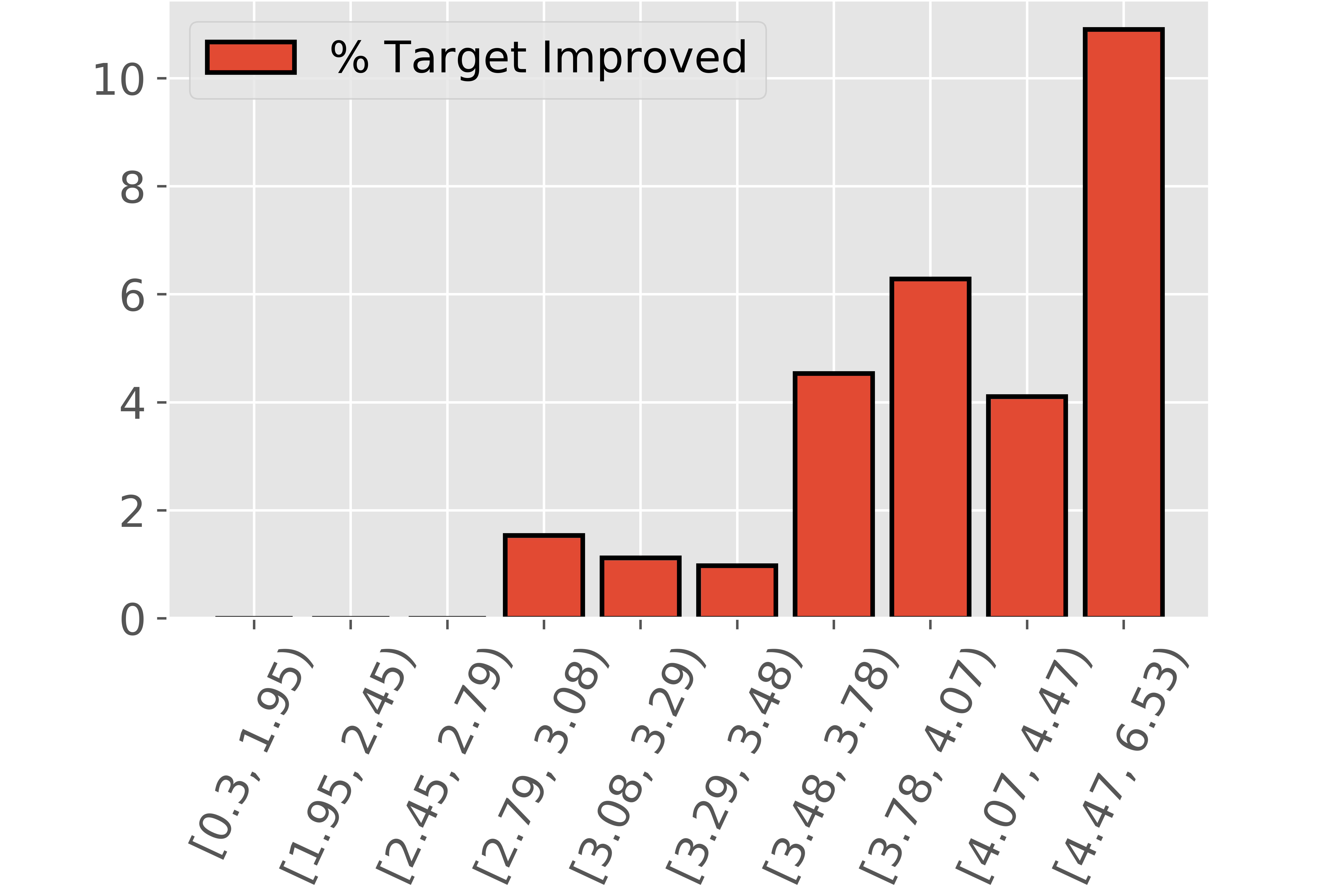}
\qquad
\includegraphics[width=0.45\textwidth]{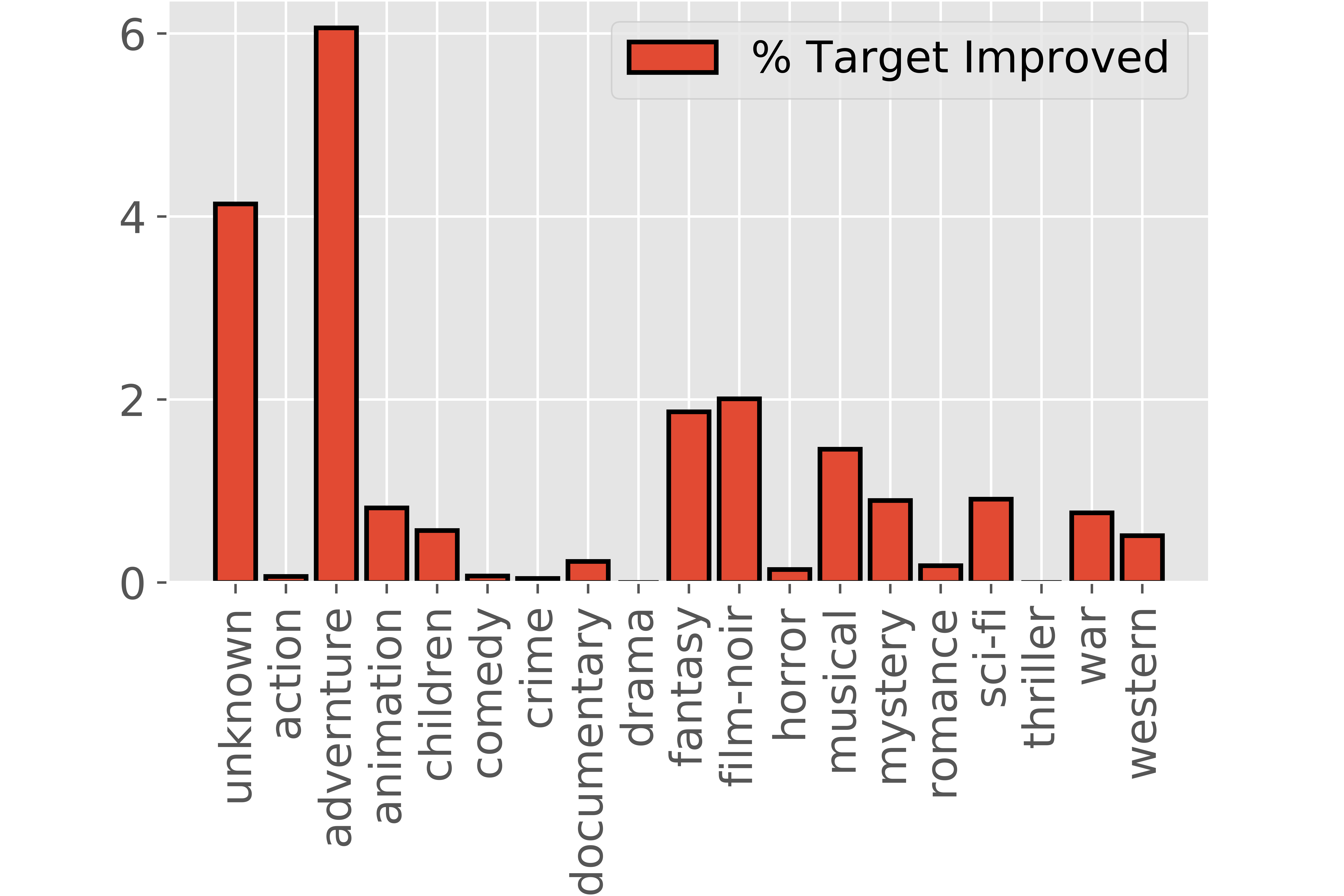}
\caption[The adversary $A$ targets the predicted score--goal (G1)--of a group of items defined over (user, item, score) entries in the target held-out set under the (E2) setup. ]
{The adversary $A$ targets the predicted score--goal (A1)--of a group of items defined over (user, item, score) entries in the target held-out set under the (E2) setup. y-axis: \% target improved (percentage of decrease in the $f_A$). x-axis: target group bins. \textbf{Left:} Groups are 10-percentile groups based on $R$'s predicted scores before the attack. \textbf{Right:} Groups are defined based on movie genres. \textbf{Gist:} \% improved is larger for groups with higher predicted score before the attack; up to 6.1\% decrease in predicted score happens for the ``adventure" genre. 
}
\label{fig:massive1ad}
\end{figure*}

Figure \ref{fig:massive1ad} focuses on goal (A1)--- decreasing the average predicted score from $R$ over an item group, defined as the mean of the predicted scores over the subset of the target held-out user-item entries belonging to that group. 
Figure \ref{fig:massive1ad}, \emph{left panel} shows that $A$ tends to be capable of larger \% target improved, i.e., larger \% of decrease in the  predicted score, for the movie groups with larger original predicted score; $A$ can achieve up to 10.9\%  for the bucket with predicted scores before the attack between [4.47, 6.53).   
This is interesting, as these are the entries which would be more likely to appear on users' lists, if the attack did not happen. Figure \ref{fig:massive1ad}, \emph{right panel} shows that when grouping the movies into buckets based on their genre, $A$ can achieve up to 6.1\% decrease in predicted score for the adventure genre;  the second largest \% was found for the unknown genre. 



\begin{figure*}[t]
\centering
\includegraphics[width=0.45\textwidth]{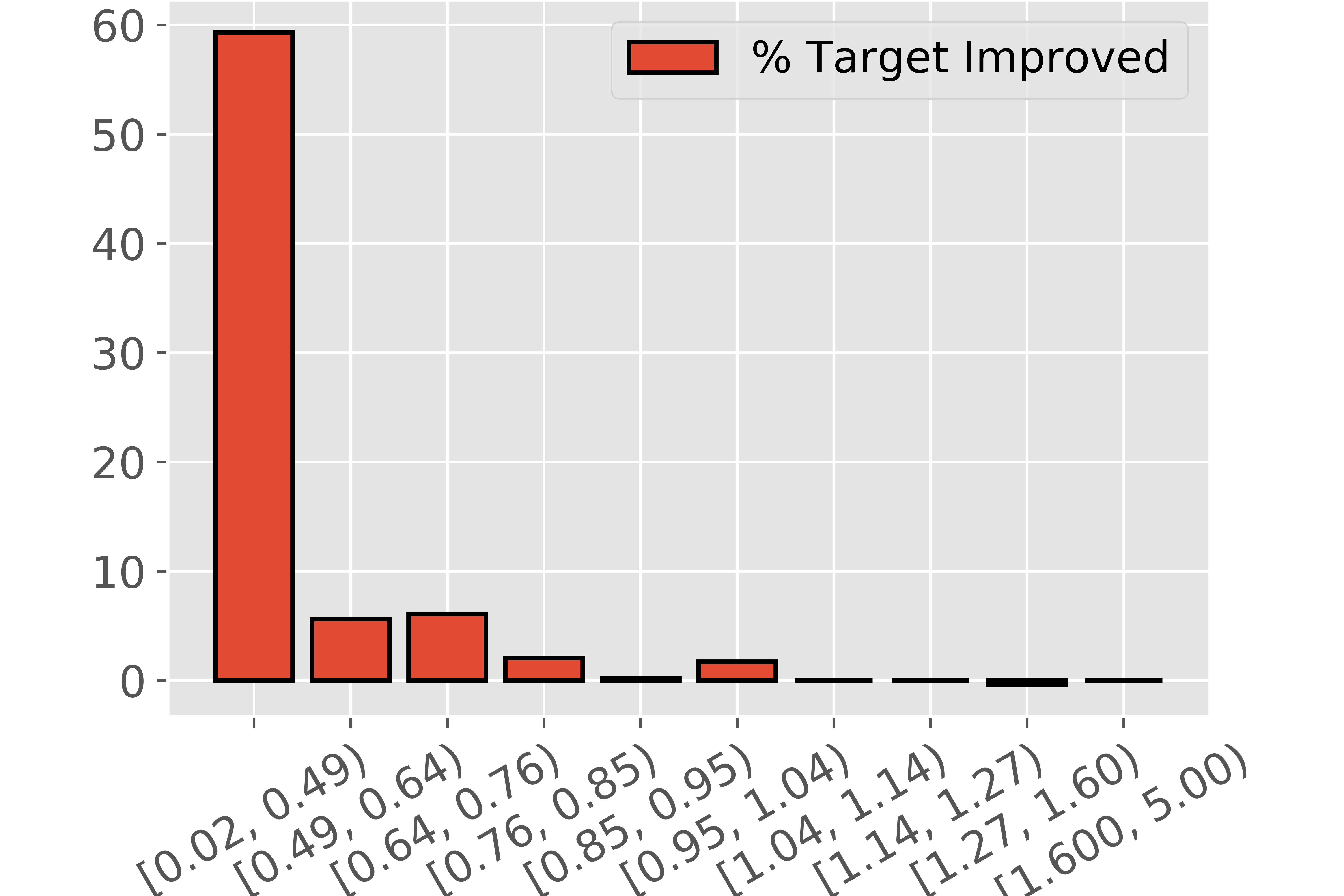}
\qquad
\includegraphics[width=0.45\textwidth]{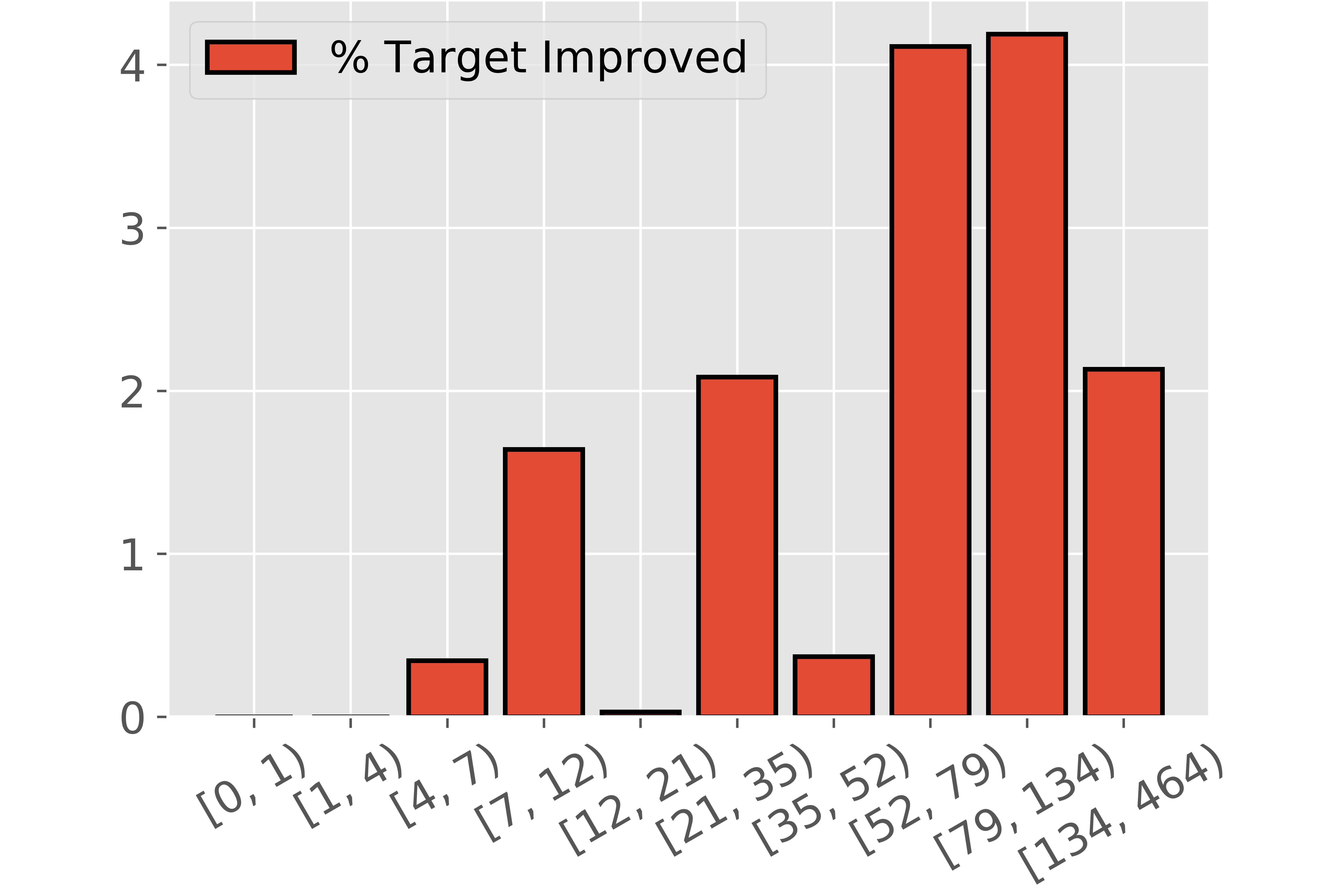}
\caption[The adversary $A$ targets the prediction error--goal (A2)--of a group of items defined over (user, item, score) entries in the target held-out set under the (E2) setup. The x-axis (groups), and y-axis (\% Improved) are as described in Figure \ref{fig:massive1ad}. ]
{
The adversary $A$ targets the prediction error--goal (G2)--of a group of items defined over (user, item, score) entries in the target held-out set under the (E2) setup. The x-axis (groups), and y-axis (\% Improved) are as described in Figure \ref{fig:massive1ad}.  \textbf{Left:} Groups are 10-percentile groups based on $R$'s initial, i.e., before the attack, prediction error. \textbf{Right:} Groups are 10-percentile groups based on number of training ratings per item. \textbf{Gist:} \% improved is larger for groups with smaller initial predicted error (left); as number of training ratings per item increase, the attack can be more successful (right). 
}
\label{fig:massive1bc}
\end{figure*}
Figure \ref{fig:massive1bc} focuses on goal (A2)---maximizing the target prediction error of a group. The \emph{left panel} shows the results for grouping movies based on the target prediction error of $R$ before the attack, and shows that $A$ can achieve up to 59.3\% target prediction error increase for the well-modeled buckets, i.e., those with [0.02, 0.49) error before the attack. The \emph{right panel} shows that $A$ can achieve up to 4.2 \% error increase  for the item bucket with [79, 134) training ratings; however, the groups with fewer than 4 training ratings were not successfully targeted. 

\subsection{Targeting Improved Modeling for a Group of Users or Items}
\label{subsec:target-improve}
In the last set of experiments, the adversary's intent is to achieve improvement in the modeling of groups of users or items in the target set---we will refer to those as \emph{targeted improvements}.  The difference between this experiment and  Section \ref{subsec:groupbased} is that here, $A$ wants to \emph{improve} how $R$ models groups of users or items, whereas before in the (A2) goal $A$ wanted to \emph{deteriorate} how $R$ modeled the groups of items. 

We examine three goals: 
\begin{itemize}
\item[(I1)] improve the average recommendation quality, as measured by Hit Rate@10 (checking whether on average per user, the user's held-out entry is included in their top-10 recommender list (hit=1), or not (hit=0)), over a user group, 
\item[(I2)] improve the modeling, as measured by mean absolute predicted error, over a group of items,  and
\item[(I3)] ensure that two user groups are equally well modeled, i.e., the gap between their modeling errors is reduced. 
\end{itemize}
Goal (I2) is essentially the negative of goal (A2). In all three (I1), (I2), (I3) goals, the metrics are defined in the target set. 

The setup followed is again the (E2) experimental setup (Section \ref{subsec:exp-design}), and the parameter setting stays the same as described in Section \ref{subsec:groupbased}. However, for the sub-experiment focusing on the (I1) goal, the target set is formed based on the (E2-a) leave-one-out setup so to compute the Hit Rates, whereas for (I2), (I3) the (E2-b) 80-10-10 split was used. 

We explore different ways of defining the target user or item groups: For the experiment realizing goal (I1) we group users into 10-percentile groups based on number of training ratings per user, or based on the side information of age. For the experiment realizing goal (I2), we create the item groups by grouping movies into 10-percentile groups based on number of training ratings per item.  Finally, for the goal (I3)-experiment, we create user groups either based on side information, dividing into two user groups based on gender, i.e., male and female, or dividing them into four 25-percentile groups using the number of training ratings per user. 

For the (I1), (I2) set of experiments, we report again the \% target improved metric, i.e., the percentage of improvement in the average Hit Rate over the users belonging to the group, or the percentage of decrease in the mean absolute error over the target set. For (I3), we measure the gap, i.e., absolute difference, among the two prediction errors in the target set we optimize $f_A$ over. 


\begin{figure*}[t]
\centering
\includegraphics[width=0.45\textwidth]{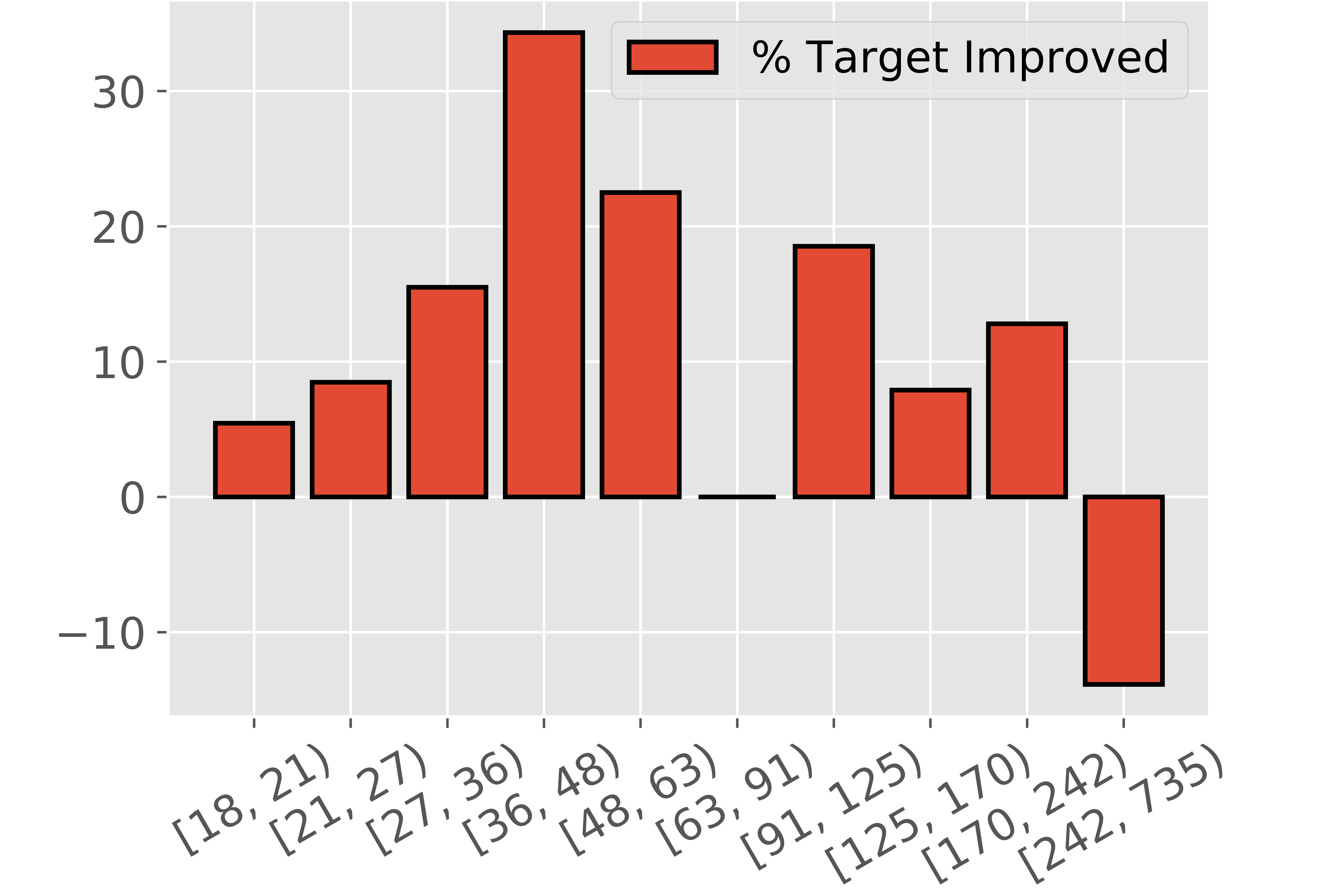}
\qquad
\includegraphics[width=0.45\textwidth]{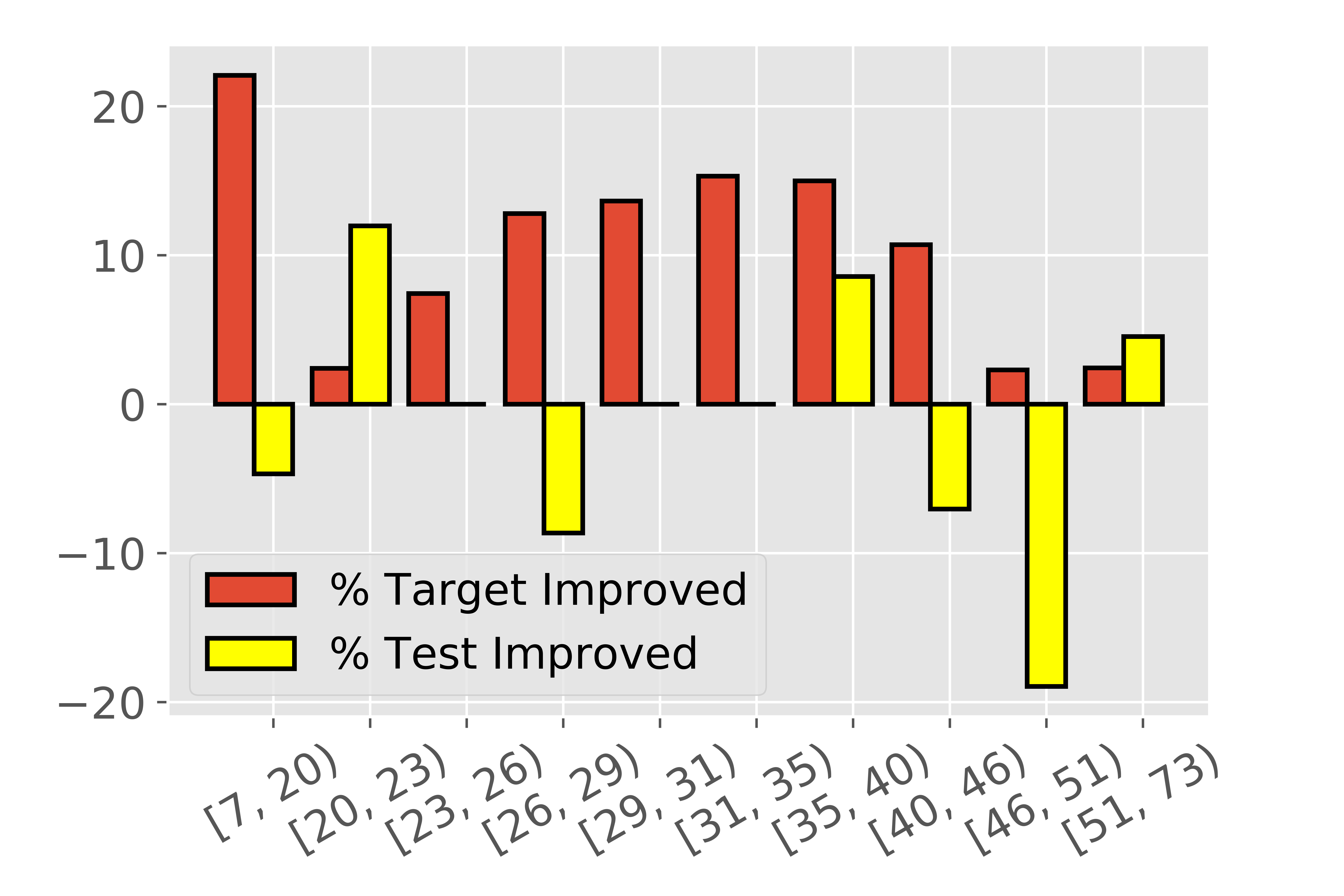}
\caption[The adversary $A$ targets to improve the recommendations, as measured by Hit Rate, of groups of users---goal (I1)---, defined over over (user, item, score) entries in the target set under the (E2) setup.]
{The adversary $A$ targets to improve the recommendations, as measured by Hit Rate, of groups of users---goal (I1)---, defined over over (user, item, score) entries in the target set under the (E2) setup. The x-axis (groups), and y-axis (\% Improved) are as described in Figure \ref{fig:massive1ad}.  \textbf{Left:} Groups are 10-percentile groups of users based on number of training ratings per user. \textbf{Right:} Groups are 10-percentile groups based on age. \textbf{Gist:} The largest improvement is observed for the [36, 48) training ratings bucket (left);  for the youngest age bucket (right). 
}
\label{fig:good1abc-a}
\end{figure*}

\begin{figure*}[t]
\centering
\includegraphics[width=0.45\textwidth]{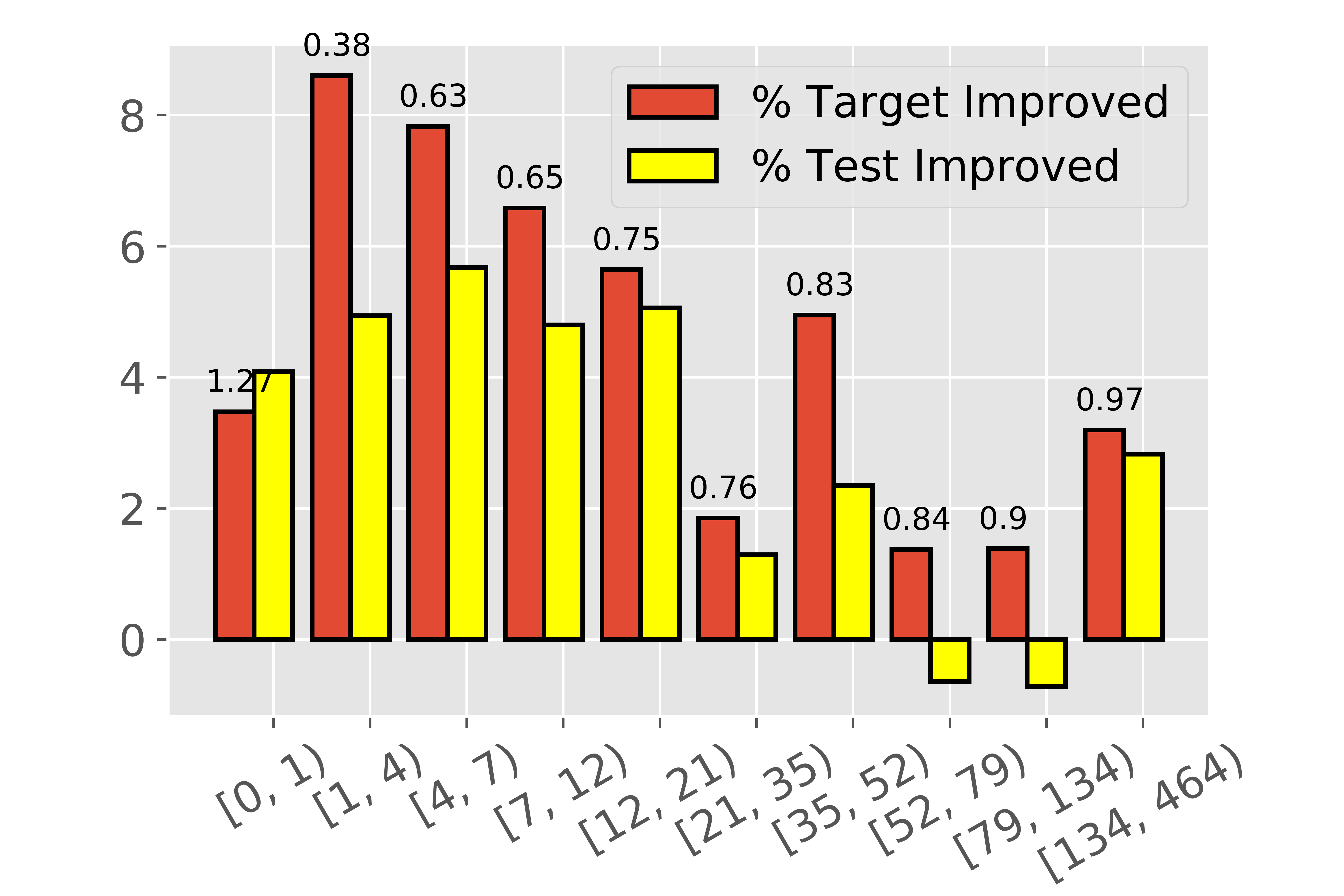}
\qquad
\rotatebox{90}{\tiny{\bf Prediction Error / \# Z-SGD Iters}}
\includegraphics[width=0.45\textwidth, height=0.28\linewidth]{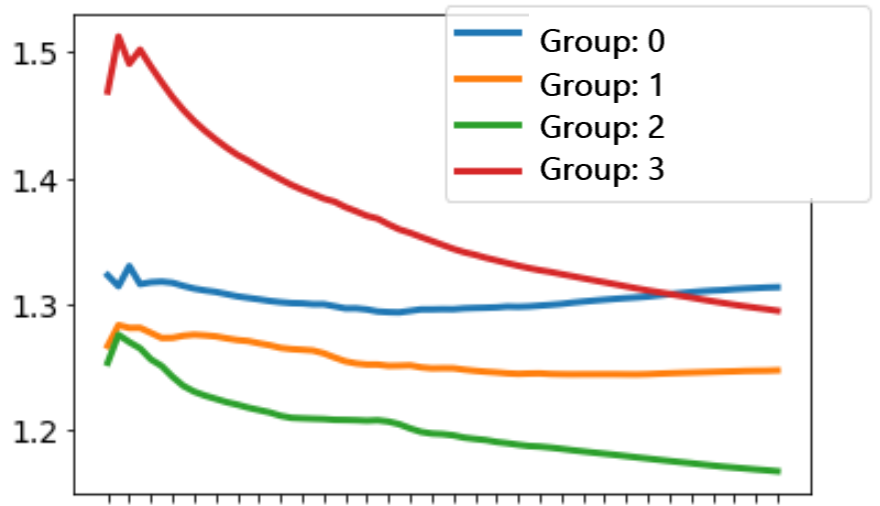}
\caption[The adversary $A$ targets to improve the modeling of groups of users or items, focusing on goals (I2), (I3). ]
{The adversary $A$ targets to improve the modeling of groups of users or items; setup is the same as in Figure \ref{fig:good1abc-a}, focusing on goals (I2), (I3).  \textbf{Left:} Goal (I2)---decrease the prediction error for groups of items, where groups are 10-percentile buckets defined based on number of training ratings per item. The x-axis shows the groups, and the annotations on top of the bars are the initial (before the attack) target prediction errors.   \textbf{Right:} Goal (I3)--- minimize the gap betwen the prediction errors of user groups, thus improving the fairness in the modeling of them. User groups are 25-percentile buckets based on number of training ratings per user. Here, the target is to reduce the gap between groups 0 and 3. The figure shows how the prediction error of each group (y-axis) varies as Z-SGD iterations (x-axis) proceed.  \textbf{Gist:} The fake users can improve the prediction error more for the item groups with smaller initial prediction error (left);  the target gap between groups 0 (blue line) and 3 (red line) becomes 0, affecting also the target prediction errors over the other groups (right).
}
\label{fig:good1abc-b}
\end{figure*}

\begin{figure*}[t]
\centering
\includegraphics[width=0.45\textwidth]{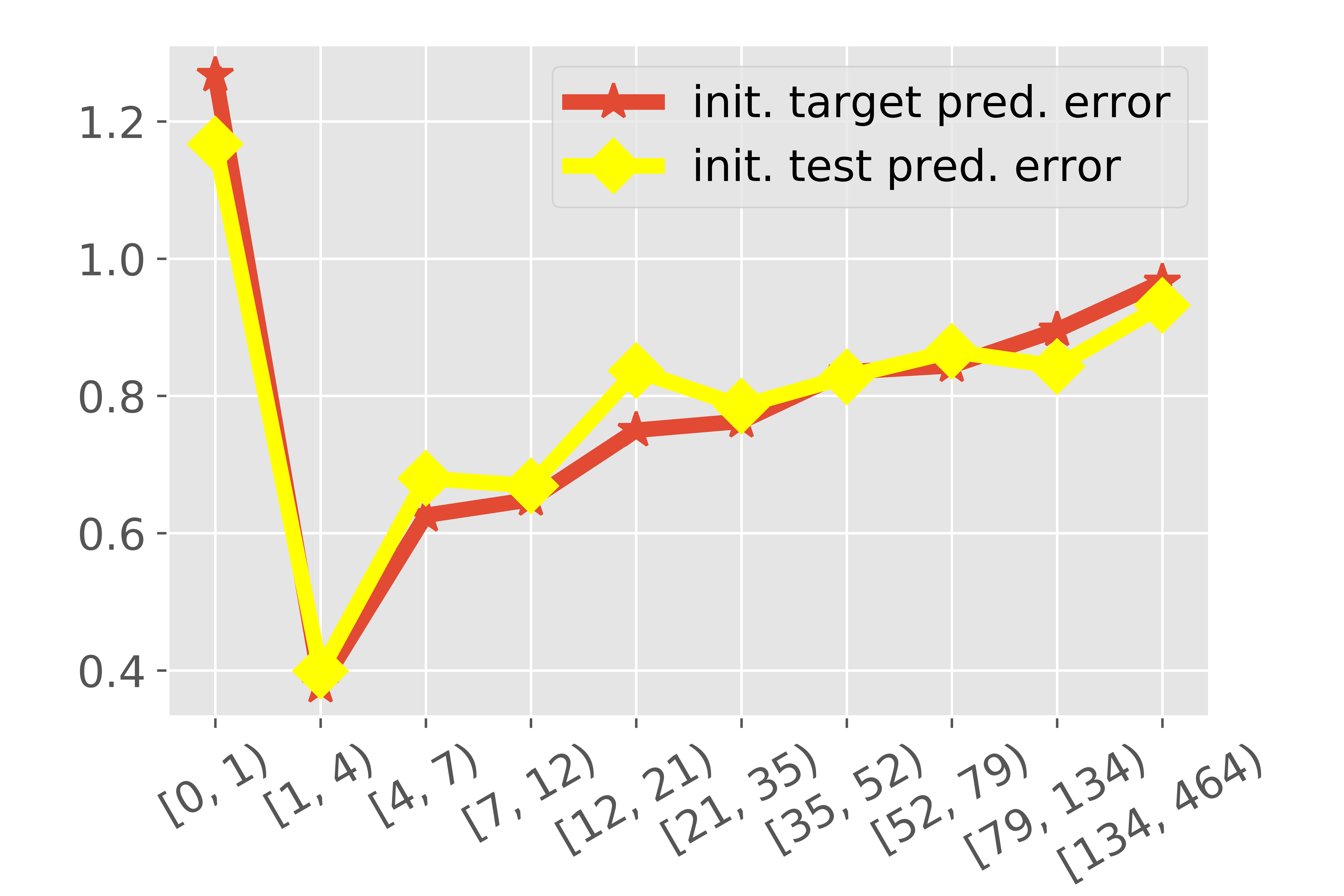}
\qquad
\includegraphics[width=0.45\textwidth]{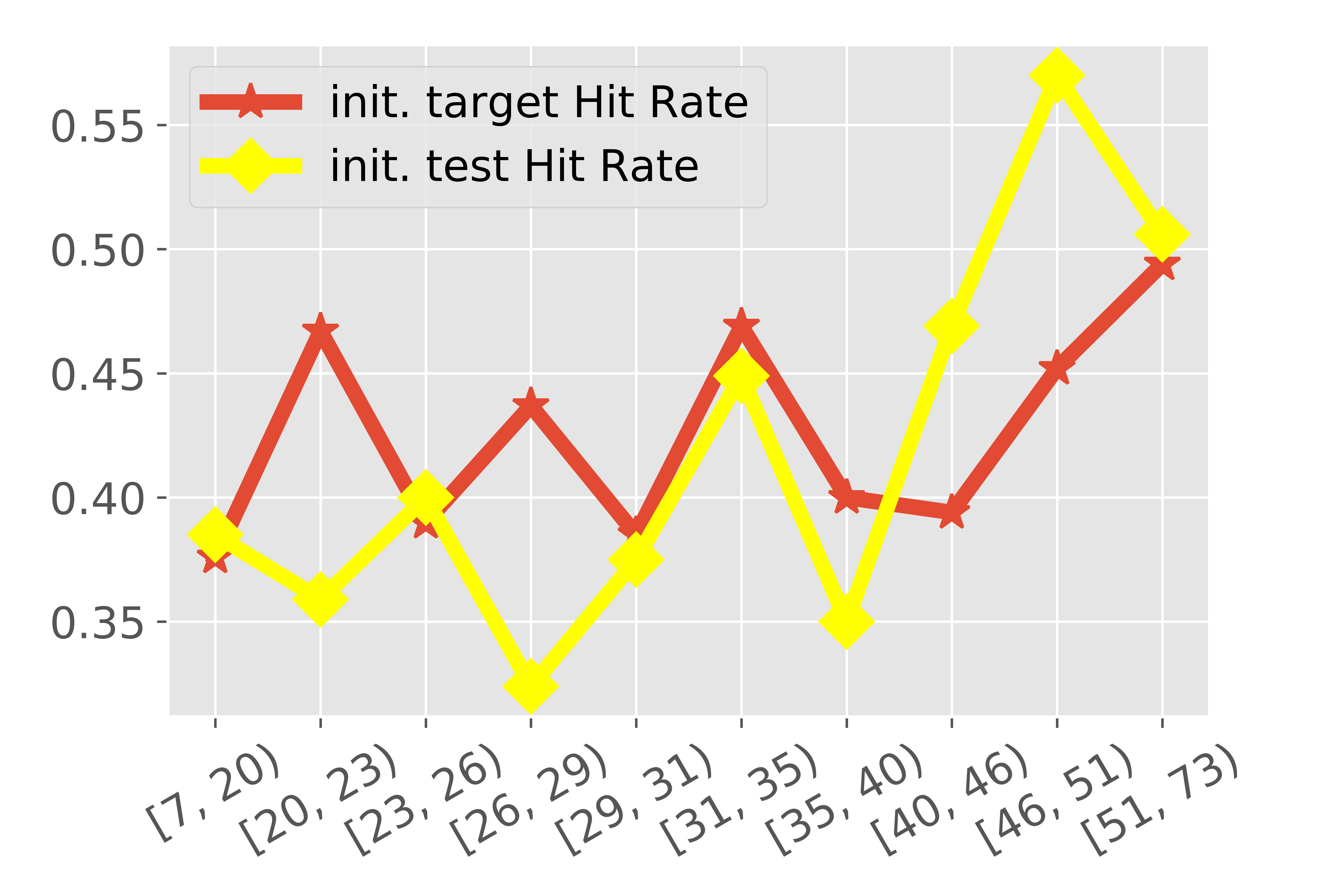}
\caption
[\textbf{Left:} Initial target and test prediction errors over ratings-based item groups.
 \textbf{Right:} Initial target and test Hit Rates over age-based user groups. ]
{\textbf{Left:} Initial target and test prediction errors over ratings-based item groups.
 \textbf{Right:} Initial target and test Hit Rates over age-based user groups. \textbf{Gist:} When the target - test trends are similar (left), the adversary tends to successfully target unseen test groups (Figure \ref{fig:good1abc-b}, \emph{left}, yellow bars); else when the trends of the initial metrics are different (right), the \% Improved trends over target and test tend to be different too (Figure \ref{fig:good1abc-a}, \emph{right}, yellow bars).  
}
\label{fig:good1abc-c}
\end{figure*}

Figure \ref{fig:good1abc-a} focuses on goal (I1), i.e., improving the Hit Rates of certain user groups. From Figure \ref{fig:good1abc-a}, \emph{left panel} we find that the user group with the largest number of ratings is not improved, in fact, it is hurt by the fake users. The groups which benefit the most from $A$ tend to be those in the middle of the rating distribution. 
 Figure \ref{fig:good1abc-a}, \emph{right panel} shows that grouping users into deciles [7, 20), [20, 23) up to [51, 73), a targeted improvement is possible; with the largest observed for the youngest age bucket. 
But, these targeted improvements do not transfer to an unseen test set from the same age group (yellow bars). We argue that this happens as the before-the-attack trends of HRs over the age groups in the target and test set differ (Figure \ref{fig:good1abc-c}, \emph{right}). 

Figure \ref{fig:good1abc-b}, \emph{left panel} focuses on goal (I2), and shows the \% improvement results in the modeling of item groups defined based on the number of training ratings per item. 
 We find that groups with the smallest target prediction errors before the ``attack" (data labels annotated on top of the bars in the plot), 
are the ones which are better targeted, i.e., the ones with the largest \% target improved. 
From Figure \ref{fig:good1abc-c}, \emph{left} we can see that the original metrics of the target and test set  hold similar trends across groups, which might be one reason why the attack generalizes here (further future analysis is needed). 

Last, we focus on goal (I3), which can be viewed as ensuring fair treatment of two user groups. By grouping users into two groups, males and females, and measuring the target mean absolute prediction error, we find that: the error for (females, males) was before the ``attack" (1.299, 1.339); during the ``attack", $A$ attempts to minimize the error of the group with the larger original error; and after the ``attack" the rounded prediction errors became equal: (1.281, 1.281), improving the absolute gap between the two groups from 0.04 to 0.0001 (without the rounding). Similar trends can be found when defining user groups based on  number of ratings, or the age side information. For example, Figure \ref{fig:good1abc-b}, \emph{right panel}  shows that for 25-percentile groups 0, 1, 2, and 3 of users based on number of training ratings per user,  the gap between group 0 and 3 becomes 0. Also, the plot shows how targeting 0-3 gap affects the prediction errors of the other groups as well as $Z$-SGD updates progress---this observation holds for all group-based attacks: targeting a group has an effect of improvement/ decrease in the other groups, too. This inter-connection among the groups seems to play a role for understanding when/ why attacks generalize to a test set (to be explored in the future).

These results serve as proof of concept that:
Together the results of this and the previous section serve as proof of concept that:
\begin{quote}
\emph{The fake users of $A$ can affect (improve/ deteriorate) how $R$ models user or item groups in a target set.}
\end{quote}

More rigorous analysis though is needed to understand patterns for the user/ item groups that can enjoy larger targeted improvements. 

\subsection*{Discussion}
Overall, our experiments indicate that an adversary can achieve a variery of intents on target groups of users or items defined in a variety of ways, while the distance between the real user and fake user distribution remains close\footnote{Although from Section \ref{subsubsec:impact} onwards we omitted the distance metrics, in all experiments mean TVD and mean JS Divergence remain close to 0.}. It is important to emphasize again that this happens under the assumptions that the adversary $A$ has the ability to: 
\begin{itemize}
\item query the recommender $R$ for predicted scores, 
\item know the underlying true distribution of user-item ratings (so to be able to create realistic-looking user profiles), and
\item under the (E2) setup, to target true---not unrated as in the (E1) setup---user-item entries; which however, might not be realizable in real-world settings,
\end{itemize}
while $R$ is oblivious to $A$, and is periodically (at every Z-SGD update) warm-start retrained over both real and fake ratings for a few alt-min iterations. 
Nevertheless, it is still notable that the adversary $A$ can affect the recommender's predictions; with perhaps the most interesting result being that just by targeting the top user predicted for an item, all top users predicted by $R$ for this item, and all top items predicted by $R$ for this user are successfully targeted as well.

\section{Related Work}
Most works on attacking a recommender system $R$ injecting fake users (i.e., ``shilling attacks"), have focused on engineering user profiles with high/ small score on the target item(s) and average or normal-distributed scores for (a subset of) the other items. These approaches vary in terms of: the {recommender model under attack} (e.g. user/ item-based collaborative filtering, model/ memory-based), the {adversary's knowledge} (e.g. of the true rating distribution, the recommender's architecture or parameters), the {attack intent} (e.g. promote/ demote an item or group), and the adversary's  {success metric} \cite{aggarwal2016attack,burke2015robust,mobasher2007toward,o2002promoting,o2004collaborative}. Our work is the first where the adversarial fake user generator $A$ \emph{learns} to generate fake profiles in an \emph{end-to-end fashion}, capturing different adversarial intents. Our approach is demonstrated for a low rank $R$ \cite{mnih2008probabilistic}, assuming that $A$ knows the true rating distribution, and can evaluate $R$'s objective, but not its gradient.


By learning fake user profiles, we attempt to bridge the gap between shilling attacks and works on adversarial examples \cite{su2017one,moosavi2016deepfool,moosavi2016universal,goodfellow2016cleverhans,goodfellow2014explaining}. 
These have largely focused on \emph{classification}; only recently a new adversarial attack type was introduced for graph data \cite{zugner2018adversarial}.
Our approach, although related with works on adversarial examples, has important differences: it  (i) consists of injecting fake user profiles during training, instead of perturbing the features of examples on a deployed trained model, and (ii) considers the \emph{recommendation} problem, which can lead to attacks of considerably bigger size than e.g. one-pixel attacks \cite{su2017one}, or small-norm perturbation attacks---especially since recommendation models rely on the assumption that similar users tend to like items similarly. 
In \cite{dalvi2004adversarial} adversarial classification was formulated in a game-theory framework, giving an optimal classifier given the adversary's optimal strategy. Our work focuses on recommendation and presents the strategy from the adversary's view, considering an oblivious recommender. 

Our work is the first to apply generative adversarial nets (GANs) \cite{goodfellow2014generative}, to produce realistic-looking fake users. Previous works have used GANs in recommenders, either for better matrix reconstruction \cite{wang2017irgan}, or to find visually similar item recommendations \cite{kang2017visually}---but never to learn the rating distribution.  


When injecting fake user profiles, the original rating data is augmented. Hence, data augmentation works become related \cite{antoniou2017data}. Also, our experiments in Section \ref{subsec:target-improve} make works 
on adversarial training to tailor representations \cite{ganin2016domain,edwards2015censoring,beutel2017data}, or on better subset modeling \cite{beutel2017beyond, christakopoulou2016local, christakopoulou2018latent} relevant.  
  

Our work does \emph{not} focus on the recommender's strategy to resist; stategies in \cite{papernot2016towards,goodfellow2018making} and robust models \cite{burke2015robust,mobasher2007toward} remain to be tested. Another direction is to relate our results to  stability \cite{adomavicius2012stability} and ratings' influence \cite{rashid2005influence,resnick2008information}.

\label{sec:adv-related}

\section{Conclusions and Future Directions}
In this paper, we presented the first work on \emph{machine learned} adversarial attacks to recommendation systems. We introduced  the framework of adversarial recommendation, posed as a game between a low rank recommender $R$ oblivious to the adversary's existence, and an adversary $A$ aiming to generate fake user profiles that are realistic-looking, and optimize some adversarial intent. Our experiments showed that adversarial attacks for a variety of intents are possible, while remaining unnoticeable. A notable attack example is that to ruin the predicted scores of a specific item for users who would have loved or hated that item, it suffices to minimize the predicted score of the top predicted user for that item before the attack. 

This study provides several interesting directions for future research on adversarial recommendation. First, our approach needs to be tested in other recommendation datasets, and against other recommendation models, while also varying the knowledge of the adversary $A$. 
Second, our work has only scratched the surface on how data augmentation with adversarially learned users can improve the modeling of certain groups; further research is needed to develop a more thorough understanding as to when it could help, and how it would compare with alternative techniques  \cite{beutel2017beyond,beutel2017data}. Third, other optimization objectives could be encoded for the adversarial intent so to craft the recommendations / representations for certain goals \cite{ganin2016domain}. Also, generating new realistic-looking user profiles could be used to improve testbeds of recommendation algorithms. Finally, one of the most important directions is to to create adversary-aware recommenders, and evaluate the degree to which they, as well as existing robust recommenders \cite{burke2015robust}, can resist to machine learned attacks.

\label{sec:adv-concl}

\vspace*{3mm}
{\bf Acknowledgements:} The research was supported by NSF grants IIS-1563950, IIS1447566,
IIS-1447574, IIS-1422557, CCF-1451986, CNS-1314560,
IIS-0953274, IIS-1029711, NASA grant NNX-12AQ39A, and gifts
from Adobe, IBM, and Yahoo. 



\begin{thebibliography}{10}

\bibitem{adomavicius2012stability}
Gediminas Adomavicius and Jingjing Zhang.
\newblock Stability of recommendation algorithms.
\newblock {\em ACM Transactions on Information Systems (TOIS)}, 30(4):23, 2012.

\bibitem{agarwal2010optimal}
Alekh Agarwal, Ofer Dekel, and Lin Xiao.
\newblock Optimal algorithms for online convex optimization with multi-point
  bandit feedback.
\newblock In {\em COLT}, pages 28--40. Citeseer, 2010.

\bibitem{aggarwal2016attack}
Charu~C Aggarwal.
\newblock Attack-resistant recommender systems.
\newblock In {\em Recommender Systems}, pages 385--410. Springer, 2016.

\bibitem{antoniou2017data}
Antreas Antoniou, Amos Storkey, and Harrison Edwards.
\newblock Data augmentation generative adversarial networks.
\newblock {\em arXiv preprint arXiv:1711.04340}, 2017.

\bibitem{beutel2017data}
Alex Beutel, Jilin Chen, Zhe Zhao, and Ed~H Chi.
\newblock Data decisions and theoretical implications when adversarially
  learning fair representations.
\newblock {\em arXiv preprint arXiv:1707.00075}, 2017.

\bibitem{beutel2017beyond}
Alex Beutel, Ed~H Chi, Zhiyuan Cheng, Hubert Pham, and John Anderson.
\newblock Beyond globally optimal: Focused learning for improved
  recommendations.
\newblock In {\em Proceedings of the 26th International Conference on World
  Wide Web}, pages 203--212. International World Wide Web Conferences Steering
  Committee, 2017.

\bibitem{bhatnagar2012stochastic}
Shalabh Bhatnagar, HL~Prasad, and LA~Prashanth.
\newblock {\em Stochastic recursive algorithms for optimization: simultaneous
  perturbation methods}, volume 434.
\newblock Springer, 2012.

\bibitem{burke2015robust}
Robin Burke, Michael~P O�Mahony, and Neil~J Hurley.
\newblock Robust collaborative recommendation.
\newblock In {\em Recommender systems handbook}, pages 961--995. Springer,
  2015.

\bibitem{christakopoulou2016local}
Evangelia Christakopoulou and George Karypis.
\newblock Local item-item models for top-n recommendation.
\newblock In {\em Proceedings of the 10th ACM Conference on Recommender
  Systems}, pages 67--74. ACM, 2016.

\bibitem{christakopoulou2018latent}
Evangelia Christakopoulou and George Karypis.
\newblock Local latent space models for top-n recommendation.
\newblock In {\em Proceedings of the 24th ACM SIGKDD International Conference
  on Knowledge Discovery \& Data Mining}, pages 1235--1243. ACM, 2018.

\bibitem{dalvi2004adversarial}
Nilesh Dalvi, Pedro Domingos, Sumit Sanghai, Deepak Verma, et~al.
\newblock Adversarial classification.
\newblock In {\em Proceedings of the tenth ACM SIGKDD international conference
  on Knowledge discovery and data mining}, pages 99--108. ACM, 2004.

\bibitem{duchi2015optimal}
John~C Duchi, Michael~I Jordan, Martin~J Wainwright, and Andre Wibisono.
\newblock Optimal rates for zero-order convex optimization: The power of two
  function evaluations.
\newblock {\em IEEE Transactions on Information Theory}, 61(5):2788--2806,
  2015.

\bibitem{edwards2015censoring}
Harrison Edwards and Amos Storkey.
\newblock Censoring representations with an adversary.
\newblock {\em arXiv preprint arXiv:1511.05897}, 2015.

\bibitem{ganin2016domain}
Yaroslav Ganin, Evgeniya Ustinova, Hana Ajakan, Pascal Germain, Hugo
  Larochelle, Francois Laviolette, Mario Marchand, and Victor Lempitsky.
\newblock Domain-adversarial training of neural networks.
\newblock {\em The Journal of Machine Learning Research}, 17(1):2096--2030,
  2016.

\bibitem{goodfellow2016deep}
Ian Goodfellow, Yoshua Bengio, Aaron Courville, and Yoshua Bengio.
\newblock {\em Deep learning}, volume~1.
\newblock MIT press Cambridge, 2016.

\bibitem{goodfellow2018making}
Ian Goodfellow, Patrick McDaniel, and Nicolas Papernot.
\newblock Making machine learning robust against adversarial inputs.
\newblock {\em Communications of the ACM}, 61(7):56--66, 2018.

\bibitem{goodfellow2016cleverhans}
Ian Goodfellow, Nicolas Papernot, Patrick McDaniel, R~Feinman, F~Faghri,
  A~Matyasko, K~Hambardzumyan, YL~Juang, A~Kurakin, R~Sheatsley, et~al.
\newblock cleverhans v0. 1: an adversarial machine learning library.
\newblock {\em arXiv preprint}, 2016.

\bibitem{goodfellow2014generative}
Ian Goodfellow, Jean Pouget-Abadie, Mehdi Mirza, Bing Xu, David Warde-Farley,
  Sherjil Ozair, Aaron Courville, and Yoshua Bengio.
\newblock Generative adversarial nets.
\newblock In {\em NIPS}, pages 2672--2680, 2014.

\bibitem{goodfellow2014explaining}
Ian~J Goodfellow, Jonathon Shlens, and Christian Szegedy.
\newblock Explaining and harnessing adversarial examples.
\newblock {\em arXiv preprint arXiv:1412.6572}, 2014.

\bibitem{hansen2001completely}
Nikolaus Hansen and Andreas Ostermeier.
\newblock Completely derandomized self-adaptation in evolution strategies.
\newblock {\em Evolutionary computation}, 9(2):159--195, 2001.

\bibitem{harper2016movielens}
F~Maxwell Harper and Joseph~A Konstan.
\newblock The movielens datasets: History and context.
\newblock {\em Acm transactions on interactive intelligent systems (tiis)},
  5(4):19, 2016.

\bibitem{kang2017visually}
Wang-Cheng Kang, Chen Fang, Zhaowen Wang, and Julian McAuley.
\newblock Visually-aware fashion recommendation and design with generative
  image models.
\newblock In {\em Data Mining (ICDM), 2017 IEEE International Conference on},
  pages 207--216. IEEE, 2017.

\bibitem{lam2004shilling}
Shyong~K Lam and John Riedl.
\newblock Shilling recommender systems for fun and profit.
\newblock In {\em WWW}, pages 393--402. ACM, 2004.

\bibitem{mnih2008probabilistic}
Andriy Mnih and Ruslan~R Salakhutdinov.
\newblock Probabilistic matrix factorization.
\newblock In {\em Advances in neural information processing systems}, pages
  1257--1264, 2008.

\bibitem{mobasher2007toward}
Bamshad Mobasher, Robin Burke, Runa Bhaumik, and Chad Williams.
\newblock Toward trustworthy recommender systems: An analysis of attack models
  and algorithm robustness.
\newblock {\em ACM Transactions on Internet Technology (TOIT)}, 7(4):23, 2007.

\bibitem{moosavi2016universal}
Seyed-Mohsen Moosavi-Dezfooli, Alhussein Fawzi, Omar Fawzi, and Pascal
  Frossard.
\newblock Universal adversarial perturbations.
\newblock {\em arXiv preprint arXiv:1610.08401}, 2016.

\bibitem{moosavi2016deepfool}
Seyed-Mohsen Moosavi-Dezfooli, Alhussein Fawzi, and Pascal Frossard.
\newblock Deepfool: a simple and accurate method to fool deep neural networks.
\newblock In {\em CVPR}, pages 2574--2582, 2016.

\bibitem{o2004collaborative}
Michael O'Mahony, Neil Hurley, Nicholas Kushmerick, and Gu{\'e}nol{\'e}
  Silvestre.
\newblock Collaborative recommendation: A robustness analysis.
\newblock {\em ACM Transactions on Internet Technology (TOIT)}, 4(4):344--377,
  2004.

\bibitem{o2002promoting}
Michael~P O’Mahony, Neil~J Hurley, and Guenole~CM Silvestre.
\newblock Promoting recommendations: An attack on collaborative filtering.
\newblock In {\em International Conference on Database and Expert Systems
  Applications}, pages 494--503. Springer, 2002.

\bibitem{papernot2016towards}
Nicolas Papernot, Patrick McDaniel, Arunesh Sinha, and Michael Wellman.
\newblock Towards the science of security and privacy in machine learning.
\newblock {\em arXiv preprint arXiv:1611.03814}, 2016.

\bibitem{radford2015unsupervised}
Alec Radford, Luke Metz, and Soumith Chintala.
\newblock Unsupervised representation learning with deep convolutional
  generative adversarial networks.
\newblock {\em arXiv preprint arXiv:1511.06434}, 2015.

\bibitem{rashid2005influence}
Al~Mamunur Rashid, George Karypis, and John Riedl.
\newblock Influence in ratings-based recommender systems: An
  algorithm-independent approach.
\newblock In {\em Proceedings of the 2005 SIAM International Conference on Data
  Mining}, pages 556--560. SIAM, 2005.

\bibitem{resnick2008information}
Paul Resnick and Rahul Sami.
\newblock The information cost of manipulation-resistance in recommender
  systems.
\newblock In {\em Proceedings of the 2008 ACM conference on Recommender
  systems}, pages 147--154. ACM, 2008.

\bibitem{su2017one}
Jiawei Su, Danilo~Vasconcellos Vargas, and Sakurai Kouichi.
\newblock One pixel attack for fooling deep neural networks.
\newblock {\em arXiv preprint arXiv:1710.08864}, 2017.

\bibitem{szegedy2013intriguing}
Christian Szegedy, Wojciech Zaremba, Ilya Sutskever, Joan Bruna, Dumitru Erhan,
  Ian Goodfellow, and Rob Fergus.
\newblock Intriguing properties of neural networks.
\newblock {\em arXiv preprint arXiv:1312.6199}, 2013.

\bibitem{wang2017irgan}
Jun Wang, Lantao Yu, Weinan Zhang, Yu~Gong, Yinghui Xu, Benyou Wang, Peng
  Zhang, and Dell Zhang.
\newblock Irgan: A minimax game for unifying generative and discriminative
  information retrieval models.
\newblock In {\em Proceedings of the 40th International ACM SIGIR conference on
  Research and Development in Information Retrieval}, pages 515--524. ACM,
  2017.

\bibitem{zugner2018adversarial}
Daniel Z{\"u}gner, Amir Akbarnejad, and Stephan G{\"u}nnemann.
\newblock Adversarial attacks on neural networks for graph data.
\newblock In {\em KDD}, pages 2847--2856. ACM, 2018.

\end{thebibliography}

\end{document}